\documentclass{ptephy_v1}
\usepackage{amsmath} 
\begin{document}
\title{
Separation of equilibrium part from an off-equilibrium state produced by 
relativistic heavy ion collisions using a scalar dissipative strength
}
\author{\name{Takeshi Osada}{} 
}

\address{\affil{}{
Department of Physics, 
Faculty of Liberal Arts and Sciences,\\ 
Tokyo City University, 
Tamazutsumi 1-28-1, Setagaya-ku, Tokyo 158-8557, Japan
}
\email{osada@ph.ns.tcu.ac.jp}}

\begin{abstract}%
We have proposed a novel way to specify the initial conditions of a 
dissipative fluid dynamical model 
for a given energy density $\varepsilon=u_{\mu}T^{\mu\nu}u_{\nu}$ 
and baryon number density $n=N^{\mu}u_{\mu}$, 
which does not impose the so-called 
 Landau matching condition for an off-equilibrium state. 
In addition to usual 
two parameters for equilibrium part, i.e., 
$\alpha\equiv \mu/T$, $\beta\equiv1/T$ (where 
$T$ is separation temperature and $\mu$ is separation chemical potential 
introduced to separate equilibrium part from the 
off-equilibrium state), a dissipative strength $\gamma$ 
is newly introduced to specify the off-equilibrium state. 
These $\alpha$, $\beta$ and $\gamma$ can be 
uniquely determined by $\varepsilon$, 
$n$ and $P_{\rm eq}(\alpha,\beta)+\Pi=-1/3\Delta_{\mu\nu}T^{\mu\nu}$ 
consisting of both kinetic theoretical definitions 
and the thermodynamical stability condition. 
For $\gamma <10^{-3}$, $T$ and $\mu$ are almost independent of $\gamma$, which means that
the Landau matching condition is approximately satisfied.  
However, this is not the case for $\gamma \gtrsim 10^{-3}$.  
\end{abstract}
\subjectindex{}
\maketitle
\section{Introduction}\label{sec:1} 
Relativistic hydrodynamical models have been applied to studies of matter that is produced
in high-energy hadron or nuclear collisions. 
Fluid dynamical descriptions, in particular, provide 
a simple picture of the space-time evolution of the hot/dense matter 
produced by ultra-relativistic heavy-ion collisions at 
RHIC and LHC \cite{Heinz2013_1, Heinz2013_2}. 
It is expected that this simple picture makes it possible to investigate 
the strongly interacting quark and gluon matter present at the initial stage of 
the collisions. 
The fluid model assumes that there exist local 
quantities of the matter such as energy density, pressure and so on. 
In particular, it is considered that the pressure gradients of the matter 
causes collective phenomena \cite{Gustafsson1984, Hirano:2008hy}. 
These expected phenomena have been successfully observed as an elliptic flow 
coefficient $v_2$ in the CERN SPS experiment NA49 \cite{NA491998}, 
at RHIC experiment \cite{PhysRevC.77.054901,PhysRevLett.98.162301,
PhysRevLett.98.242302} and 
recent ALICE experiments in LHC \cite{Snellings2014} including the
higher order flow harmonic $v_n$ for ($n=3,4$). These have been observed 
as a function of various characteristics, including the 
transverse momentum $p_{\rm T}$ or rapidity $y$ and so on. 
Hence, the hydrodynamical model 
has been widely accepted 
because of such experimental evidences.
To investigate the properties of the quark and gluon matter 
created during such ultra-relativistic heavy-ion collisions more precisely, 
it is necessary to consider the effects of the viscosities and corresponding dissipation
\cite{MurongaPRC69}. 
These effects are introduced into the hydrodynamic 
simulations and a detailed comparison between simulation and 
experimental data is made 
(see, for example, Ref.\cite{SongPRC89}).   
However, as several authors have noted, dissipative hydrodynamics is not 
yet completely understood and there are issues associated to 
the determination of the hydrodynamical flow 
\cite{TsumuraPLB646,TsumuraPLB690, Van2013} (see also, Ref. \cite{Arnold2014}). 
In this article, the Landau matching (fitting) condition that is necessary to specify the
initial conditions for 
dissipative fluid dynamics is discussed. This may be related to the issue 
of defining the local rest frame \cite{TsumuraPLB690}. 

The fundamental equations of relativistic 
fluid dynamics are defined by the conservation laws of 
 energy-momentum and the charge (in this paper, 
we assume the net baryon density as the conserved charge), 
\begin{subequations}
\begin{eqnarray}
   \partial_{\mu} T^{\mu\nu}(x) =0, \\
   \partial_{\mu} N^{\nu}(x)=0. 
\end{eqnarray}
\end{subequations}
Here, $T^{\mu\nu}$ and $N^{\mu}$ are respectively the 
energy-momentum tensor and the
conserved charge current at a given point in space-time $x$, 
which can be obtained by a 
coarse-graining procedure \cite{KodamaEPJA48} with some finite size 
(fluid cell size), $l_{\rm fluid}$. 
Hence, the fluid dynamical model expressed
as a coarse-graining theory 
describing macroscopic phenomena 
can be 
derived from the underlying kinetic theory.
In the case of a perfect fluid limit, 
the microscopic collision time scale $\tau_{\rm micro}$ is 
 much shorter than the macroscopic 
evolution time scale $\tau_{\rm macro}$\cite{PhysRevC.73.034904}, 
thus 
\begin{eqnarray}
  \tau_{\rm macro} \gg \tau_{\rm micro}. 
  \label{eq:perfect_condition}
\end{eqnarray}
If the condition in eq.(\ref{eq:perfect_condition}) is satisfied, 
the distribution function 
instantaneously relaxes to its local equilibrium form. 
In the local rest frame of the fluid, i.e., the frame in which 
the fluid velocity is given by $u^{\mu}(x)=(1,0,0,0)$, 
the local equilibrium distribution functions 
for particles and for anti-particles are respectively 
given (within the Boltzmann approximation) as
\begin{subequations}\label{eq:equilibrium_distribution}
\begin{eqnarray}
  f_{{\bf k}0}(x)= \exp[\frac{-k_{\mu}u^{\mu}(x)+\mu_0(x)}{T_0(x)}], \\
  \bar{f}_{{\bf k}0}(x)= \exp[\frac{-k_{\mu}u^{\mu}(x)-\mu_0(x)}{T_0(x)}],
\end{eqnarray}
\end{subequations}
where $k^{\mu}$ is a four momentum vector of a particle or an 
anti-particle within 
a cube 
the coarse-graining scale $l_{\rm fluid}$ on a side (fluid cell).  
We decompose $k^{\mu}$ by using the local flow vector $u^{\mu}$, i.e., 
$k^{\mu}=(k^{\lambda}u_{\lambda})u^{\mu}+k^{\langle \mu\rangle}$ 
where $k^{\mu}u_{\mu}\equiv E_{\bf k}$ is a scalar and 
$k^{\langle \mu\rangle}\equiv \Delta_{\nu}^{\mu} k^{\nu}$ is a 
vector, which  is orthogonal to $u^{\mu}$. 
In the local rest frame, the scalar $E_{\bf k}$ coincides with the
zero-th component of the four vector, energy of the classical particle, $k^0$. 
The projection tensor onto the 3-space that is orthogonal to the flow velocity is defined
by $\Delta^{\mu\nu} \equiv g^{\mu\nu}-u^{\mu}u^{\nu}$ with 
$g^{\mu\nu}$ being the metric tensor, $g^{\mu\nu}={\rm diag}(+1,-1,-1,-1)$. 
$T_0(x)$ and $\mu_0(x)$ are the equilibrium temperature and chemical potential 
respectively. 
When the distribution function is given by eq.(\ref{eq:equilibrium_distribution}), 
the energy-momentum tensor and the net baryon number current vector are defined as 
\begin{subequations}
\begin{eqnarray}
    T^{\mu\nu}_{\rm eq} &=&
    \int\!\!\frac{d^3{\bf k}k^{\mu}k^{\nu}}{(2\pi)^3k^0} 
    \Big[ f_{{\bf k}0}(x) + \bar{f}_{{\bf k}0}(x) \Big] \nonumber \\
&=& \varepsilon_{\rm eq} u^{\mu}u^{\nu} 
   -P_{\rm eq} (\varepsilon_{\rm eq},n_{\rm eq}) \Delta^{\mu\nu}, 
    \\ 
    N^{\mu}_{\rm eq} &=&  \int\!\!\frac{d^3{\bf k}~k^{\mu}}{(2\pi)^3k^0} 
    \Big[ f_{{\bf k}0}(x) - \bar{f}_{{\bf k}0}(x) \Big] \nonumber \\
&=& n_{\rm eq} u^{\mu}, 
\end{eqnarray}
\end{subequations}
respectively, where $\varepsilon_{\rm eq}$ is the energy density in the local rest frame, 
$n_{\rm eq}$ is the net baryon density in the local rest frame, 
and $P_{\rm eq}$ is the pressure in the equilibrium state.   

If the fluid expands very rapidly (i.e., the macroscopic evolution time 
scale $\tau_{\rm macro}$ becomes shorter), especially for
 fluids produced by ultra-relativistic heavy-ion collisions, 
 equation (\ref{eq:perfect_condition}) 
may not be satisfied everywhere in the fluid during the early stages. 
In such cases, microscopic processes cannot keep pace with 
the changes in local energy and baryon density; therefore, 
we can write the following equation: 
\begin{eqnarray}
    \tau_{\rm micro} \gtrsim \tau_{\rm macro} . 
    \label{eq:inparfect_condtion}
\end{eqnarray}
Under the condition eq.(\ref{eq:inparfect_condtion}), 
the distribution function in the local rest frame does not obey the local equilibrium 
form eq.(\ref{eq:equilibrium_distribution}) and hence 
we have
\begin{subequations} 
\label{eq:modification_in_distribution}
\begin{eqnarray}
   f_{\bf k}(x)= f_{{\bf k}0}[T(x),\mu(x)] + \delta f_{\bf k}(x), \\ 
   \bar{f}_{\bf k}(x)= \bar{f}_{{\bf k}0}[T(x),\mu(x)] + \delta \bar{f}_{\bf k}(x), 
\end{eqnarray}
\end{subequations}
where $\delta f_{\bf k}$ and $\delta \bar{f}_{\bf k}$ are the deviations from 
the corresponding equilibrium distribution functions.   
We define a temperature $T(x)$ and  
 chemical potential $\mu(x)$ (hereafter the  
{\it separation temperature} and 
{\it separation chemical potential}, respectively) 
that is distinguished from the equilibrium temperature $T_0(x)$ and 
chemical potential $\mu_0(x)$ 
because we assume that 
$\delta f_{\bf k}$ and $\delta \bar{f}_{\bf k}$ 
also contribute to the energy and 
baryon number density of the system.  
In the inside of the fluid cell, within scales less than the coarse-graining scale 
$l_{\rm fluid}$, the extremely rapid expansion of matter causes 
additional tiny disturbances to the local flow 
velocity field $u^{\mu}(x)$. 
Although such tiny disturbance in the flow vectors may be cancelled out by defining a new
local rest frame (because of the randomness of such tiny disturbances), 
it also causes a disturbance in the distribution function 
$\delta f_{\bf k}(x)$ and $\delta\bar{f}_{\bf k}(x)$. 
For simplicity, we assume that the fluid considered is stable 
against such disturbances; otherwise, 
the disturbance would escalate and the flow would 
eventually become turbulent. 
The space-time evolution of the disturbance flow is 
independent from that of main background flow 
even if heat is supplied by the main background flow.  
Since the disturbance flow has no mechanism for obtaining 
energy other than heat originated 
from the shear viscosity of the main background flow, 
it is finally dissipated as heat. 
Thus, the disturbances, $\delta f_{\bf k}$ and 
$\delta \bar{f}_{\bf k}$, do not belong to any equilibrium state.  
As the macroscopic evolution time scale 
$\tau_{\rm macro}$ grows longer due to the expansion of matter 
and the pressure gradients of the fluid decrease,  
$\delta f_{\bf k}$ and $\delta \bar{f}_{\bf k}$ approaches to zero 
(an assumption of hydrodynamic stability), 
and the condition eq.(\ref{eq:perfect_condition}) 
is restored and local equilibrium is achieved. 

The question we now consider is 
{\it how to find local separation temperature $T(x)$ and 
chemical potential $\mu(x)$} 
in eq.(\ref{eq:modification_in_distribution}) 
from a given off-equilibrium state characterized by
\begin{subequations}
\begin{eqnarray}
    T^{\mu\nu}(x) &\equiv&
    \int\!\!\frac{d^3{\bf k}k^{\mu}k^{\nu}}{(2\pi)^3k^0} 
    \Big[ f_{\bf k}(x) + \bar{f}_{\bf k}(x) \Big], 
    \label{eq:energy-momentum_tensor}\\ 
    N^{\mu} (x) &\equiv&  \int\!\!\frac{d^3{\bf k}~k^{\mu}}{(2\pi)^3k^0} 
    \Big[ f_{\bf k}(x) - \bar{f}_{\bf k}(x) \Big]. 
    \label{eq:particle_current} 
\end{eqnarray}
\end{subequations} 
Usually the local separation temperature $T(x)$ and 
chemical potential $\mu(x)$ in eq.(\ref{eq:modification_in_distribution}) 
are determined by imposing the so-called Landau matching 
conditions \cite{Landau1959,Israel118,PhysRevC.73.034904,Heinz:2009xj} 
\begin{subequations}\label{eq:normal-matching1}
\begin{eqnarray}
&& \delta T^{\mu\nu} u_{\mu} u_{\nu} =0, \label{eq:normal-matching11}\\ 
&& \delta N^{\mu} u_{\mu} =0,\label{eq:normal-matching12}
\end{eqnarray}
\end{subequations} 
where $\delta T^{\mu\nu}\equiv T^{\mu\nu}-T_{\rm eq}^{\mu\nu}$ and 
$\delta N^{\mu}\equiv N^{\nu}-N_{\rm eq}^{\mu}$ are 
 the deviation of the energy-momentum tensor and the baryon charge 
current from the {\it matching state}, respectively. 
In this procedure, it is necessary to select a Lorentz frame. 
There are usually
two natural choices for the Lorentz frame, 
namely the Landau frame \cite{Landau1959} 
and the Eckart frame \cite{Eckart1940}. 
If the Landau frame is employed, for example,  
the local flow velocity $u^{\mu}$ is determined by 
the eigenvector of the following eigenvalue equation,
\begin{subequations}
\begin{eqnarray}
  u_{\mu} T^{\mu\nu} = \varepsilon u^{\nu}, 
  \label{eq:Landau_frame}
\end{eqnarray} 
where the eigenvalue $\varepsilon$ is the energy density 
measured in the rest frame. 
In the Landau matching condition, shown in eq.(\ref{eq:normal-matching11}), 
the energy density should be matched to an energy density in an equilibrium state 
parameterized by a temperature $T$ and a chemical potential $\mu$, i.e., 
$\varepsilon=\varepsilon_{\rm eq}(T,\mu)$.  
Using the flow velocity obtained in eq.(\ref{eq:Landau_frame}), we obtain the following
equation: 
\begin{eqnarray}
  N^{\mu} = n u^{\mu} +V^{\lambda}\Delta_{\lambda}^{\mu}  
  \label{eq:current_decomposition}
\end{eqnarray}
\end{subequations}
for the net baryon number current. 
Then eq.(\ref{eq:normal-matching12}) yields 
$n=n_{\rm eq}(T,\mu)$. In this way it is possible to 
 determine $T$ and $\mu$ from
$\varepsilon=\varepsilon_{\rm eq}(T,\mu)$ and $n=n_{\rm eq}(T,\mu)$.  
However, if the Eckart frame is employed, 
a different temperature $T'$ and chemical potential $\mu'$ 
may be obtained as compared to those derived 
in the Landau frame. For a paper which deals with the issue of the frame 
in the relativistic dissipative hydrodynamical model, see, 
for example, Ref.\cite{Van2013}. 
Therefore, for an off-equilibrium states, 
the temperature and chemical potential which characterize 
the distribution function for the equilibrium part may be dependent on the frame employed. 

Note that, the temperature $T$ and 
chemical potential $\mu$ are not only dependent on 
the frame employed but also on the matching condition eq.(\ref{eq:normal-matching1}). 
However, it is not always appropriate, as discussed below. 
In order to investigate how the matching condition affects the 
separation to an equilibrium part from an off-equilibrium, one must fix the 
Lorentz frame to define the fluid velocity.
We hereafter use the Landau frame and define a flow vector 
$u^{\mu}$ as an eigenvector of eq.(\ref{eq:Landau_frame}), 
but we do not use the Landau-matching condition eq.(\ref{eq:normal-matching1}); 
i.e., the obtained eigenvalue $\varepsilon$ of eq.(\ref{eq:Landau_frame}) is not 
assumed to be same as an equilibrium energy density $\varepsilon_{\rm eq}$. 

In the kinetic approach, 
since the energy-momentum tensor and 
the conserved charge current are defined 
by the second and first moment of the single particle distribution function, 
the Landau matching conditions 
eq.(\ref{eq:normal-matching1}) are equivalent to the following expressions
\cite{BiroMolnarEPJA48}
\begin{subequations} 
\begin{eqnarray}
&& \int\!\!\frac{d^3{\bf k} ~E_k^2 }{(2\pi)^3k^0} 
    \Big[ \delta f_{\bf k}(x) + \delta \bar{f}_{\bf k}(x) \Big]=0, \\ 
&& \int\!\!\frac{d^3{\bf k}~E_k}{(2\pi)^3k^0} 
    \Big[ \delta f_{\bf k}(x) - \delta \bar{f}_{\bf k}(x) \Big]=0.  
\end{eqnarray}
\end{subequations}
Clearly, these conditions strongly constrain the disturbances 
$\delta f_{\bf k}$ and $\delta \bar f_{\bf k}$ and they may distort the distribution function unnaturally. 
To restore the physical meaning of $\delta f_{\bf k}$ and $\delta \bar{f}_{\bf k}$, 
it is necessary to exclude the restrictions imposed by 
eq.(\ref{eq:normal-matching1}) and 
to generalize the Landau matching condition the
following form. 
\begin{eqnarray}
     \delta T^{\mu\nu} u_{\mu} u_{\nu} 
     =\Lambda, \quad \delta n^{\mu} u_{\mu} =\delta n. 
     \label{eq:extended-matching}
\end{eqnarray}
Here, $\Lambda$ and $\delta n$ can be considered as the 
energy density and net baryon number density 
of the disturbance (tiny turbulent) flow caused 
by rapid expansion. 
Except for particles constituting the disturbance flow, 
it is assumed that the remaining particles in the fluid cell 
approximately obey the local thermal distribution form 
$f_{{\bf k}0}$ or $\bar{f}_{{\bf k}0}$. 
Since the disturbance flow carry finite energy and 
baryon number, the corresponding separation 
temperature $T$ and separation chemical potential $\mu$ 
of the equilibrium part should be different from those 
obtained at the equilibrium limit, i.e., $T(x) \ne T_0(x)$ 
and $\mu(x) \ne \mu_0(x)$. 
Therefore, to find the separate 
temperature $T$ and chemical potential $\mu$ 
in eq.(\ref{eq:modification_in_distribution}), it is required to find
 {\it find $T$ and $\mu$ as functions of 
$\delta f_{\bf k}$ and $\delta \bar{f}_{\bf k}$,} 
\begin{subequations}
\label{eq:setting_problem}
\begin{eqnarray}
 T&=& T(\delta f_{\bf k}, \delta \bar{f}_{\bf k};T_0,\mu_0),\\
 \mu&=&\mu(\delta f_{\bf k}, \delta \bar{f}_{\bf k};T_0,\mu_0), 
\end{eqnarray} 
\end{subequations}
with boundary conditions $T\to T_0$ and $\mu \to \mu_0$ 
when the disturbance flow disappears,  
$\delta f_{\bf k}, \delta \bar{f}_{\bf k} \to 0$. 
The boundary condition corresponds to the so-called 
Landau matching condition. 

In any off-equilibrium state, most literature using the Landau matching 
condition published to date assumes that $T\equiv T_0$ and $\mu\equiv \mu_0$. 
However, in this paper, we consider the problem where $T$ and $\mu$ are dependent on the 
{\it strength of the off-equilibrium state}, i.e., $T$ and $\mu$ are functions of 
$\delta f_{\bf k}$ and $\delta \bar{f}_{\bf k}$, 
as seen in eq.(\ref{eq:setting_problem}). 

This article is organized as follows. 
In Section \ref{sec:2}, we obtain an expression for
eq.(\ref{eq:modification_in_distribution}) 
using the irreducible tensor expansion technique for the 
off-equilibrium distribution function, as recently employed 
by G.S. Denicol and his collaborators\cite{PhysRevD.85.114047}. 
We extend the formulation by introducing the general matching 
condition of eq.(\ref{eq:extended-matching}) and 
also apply the irreducible tensor expansion technique to an off-equilibrium 
entropy current and find a condition of thermodynamical stability. 
Then, we obtain a relation between 
the quantities introduced by eq.(\ref{eq:extended-matching}) and 
 determine them as functions of $\alpha, \beta$ and $\gamma\equiv \Pi/P_{\rm eq}$.  
In Section \ref{sec:3}, we demonstrate the separation of the corresponding equilibrium
state 
from the given off-equilibrium state, i.e., we determine $T$ and $\mu$ for a given 
off-equilibrium state characterized by $T_0$, $\mu_0$,  and $\gamma$.  
We also show an off-equilibrium distribution function and discuss the effects of 
the strength of the off-equilibrium state $\gamma$. 
Finally, Section \ref{sec:4} contains conclusions and summary. 
The derivations of key equations in the relativistic kinetic theory and 
 equations related to irreducible expansion theory are presented 
in Appendixes \ref{app:A}-\ref{app:G}. 

\section{Separation equilibrium part from the off-equilibrium distribution
function}\label{sec:2}
\subsection{Expansion of the single particle distribution function by irreducible
tensors}
Consider a relativistically expanding fluid in which microscopic processes 
cannot keep pace with the quick macroscopic changes. 
Let us rewrite eq.(\ref{eq:modification_in_distribution}), 
which is expected the distribution function 
under the condition eq.(\ref{eq:inparfect_condtion}), 
as the following form: 
\begin{subequations}\label{eq:phi_functional}
\begin{eqnarray} 
   f_{\bf k}(x)= f_{{\bf k}0}(T[\phi_{\bf k},\bar{\phi}_{\bf k}],
   \mu[\phi_{\bf k},\bar{\phi}_{\bf k}])
   \left( 1+\phi_{\bf k} \right) , \\
  \bar f_{\bf k}(x)= \bar f_{{\bf k}0}(T[\phi_{\bf k},\bar{\phi}_{\bf k}],
  \mu[\phi_{\bf k},\bar{\phi}_{\bf k}])
  \left( 1+\bar\phi_{\bf k} \right),
\end{eqnarray}
\end{subequations}
where $\phi_{\bf k}(x)\equiv \delta f_{\bf k}(x)/f_{{\bf k}0}(x)$ 
and  $\bar{\phi}_{\bf k}(x) \equiv \delta \bar{f}_{\bf k}(x)/\bar{f}_{{\bf k}0}(x)$
 are deviations from the local thermal equilibrium function 
eq.(\ref{eq:equilibrium_distribution}). 
The deviations $\phi_{\bf k}(x)$ and $\bar{\phi}_{\bf k}(x)$ involve information about a
given 
off-equilibrium state characterized not only by scalars 
such as  $\Lambda$, $\delta n$, and $\gamma\equiv \Pi/P_{\rm eq}$ 
but also by vectors and tensors, for example, the heat flow 
vector $W^{\mu}$, shear tensor $\pi^{\mu\nu}$ of the dissipative fluid and so on. 
In order to expand $\phi_{\bf k}$ and $\bar{\phi}_{\bf k}$ by those fluid dynamical quantities,  
it is necessary to use the orthogonal base of irreducible tensors 
\cite{DeGroot:1980dk, PhysRevD.85.114047},  
$\{1, k_{\langle \mu \rangle}, \cdots, k_{\langle \mu_1} 
\cdots k_{\mu_l\rangle} \}$, 
where the irreducible tensor of the $l$-rank is defined by 
\begin{eqnarray}
    \label{eq:projection} 
    k_{\langle \mu_1} \cdots k_{\mu_l\rangle} \equiv 
    \Delta_{\mu_1 \cdots \mu_n}^{\nu_1\cdots\nu_n} k_{\nu_1}\cdots k_{\nu_n}, 
\end{eqnarray} 
and the projection tensor $\Delta_{\mu_1 \cdots \mu_n}^{\nu_1\cdots\nu_n}$ used 
in eq.(\ref{eq:projection}) 
are defined in Appendix \ref{app:A} 
(see also Appendix \ref{app:B}, \ref{app:C} and references 
\cite{DeGroot:1980dk, PhysRevD.85.114047}). These irreducible tensors 
$k_{\langle \mu_1} \cdots k_{\mu_l\rangle}$ satisfy the 
following orthogonal condition (for derivation, see Appendix \ref{app:B}, \ref{app:C} and \ref{app:D}), 
\begin{subequations}\label{eq:orthogonality_condition_final}
\begin{eqnarray} 
&&\int\!\!\frac{d^3{\bf k} ~F(E_{\bf k})}{(2\pi)^3k^0}~ 
  k^{\langle \mu_1} 
  \cdots k^{\mu_m\rangle} k_{\langle \nu_1}
  \cdots k_{\nu_n\rangle} 
  =\Delta^{\mu1\cdots\mu_m}_{\nu_1\cdots\nu_n} 
  \delta_{mn} ~{\cal J}_m,  
\end{eqnarray}
where
\begin{eqnarray} 
&& {\cal J}_m=\frac{m!}{(2m+1)!!}
         \int\!\! \frac{d^3{\bf k} ~F(E_{\bf k})}{(2\pi)^3k^0}
         [\Delta_{\alpha\beta}k^{\alpha}k^{\beta}]^m,
     \label{eq:orthogonality_condition_final_b}
\end{eqnarray}
\end{subequations}  
and $F(E_{\bf k})$ is an arbitrary function of the energy $E_{\bf k}$. 
Since the tensors defined in eq.(\ref{eq:projection}) are orthogonal, 
we may expand the deviation $\phi_{\bf k}$ and $\bar{\phi}_{\bf k}$ as follows:  
\begin{subequations}
\label{eq:expansion-lambda}
\begin{eqnarray}
    \phi_{\bf k}(x) \equiv \sum_{l=0}^{\infty} 
    \lambda_{\bf k}^{\langle \mu_1\cdots \mu_l \rangle} 
    k_{\langle \mu_1}\cdots k_{\mu_l \rangle}, 
    \label{eq:expansion-lambda1}\\
   \bar\phi_{\bf k}(x) \equiv \sum_{l=0}^{\infty} 
    \bar\lambda_{\bf k}^{\langle \mu_1\cdots \mu_l \rangle} 
    k_{\langle \mu_1}\cdots k_{\mu_l \rangle}, 
     \label{eq:expansion-lambda2}
\end{eqnarray}
\end{subequations}
respectively, where $\lambda_{\bf k}^{\langle \mu_1\cdots \mu_l \rangle} $ and 
$\bar\lambda_{\bf k}^{\langle \mu_1\cdots \mu_l \rangle}$ are 
coefficient tensors of $l$-rank in the above expansions.  
Note that the coefficient tensors in the above expansion have momentum dependence 
(denoted by the subscript ${\bf k}$). Therefore, 
we further expand these coefficients tensors 
according to a set of polynomial functions of the energy $E_{\bf k}$ 
having a maximum order of $n$ as used in Ref. \cite{DeGroot:1980dk},
\begin{eqnarray}
  P^{(l)}_{{\bf k}n} =\sum_{r=0}^{n} a_{nr}^{(l)} (E_{\bf k})^r~. 
  \label{eq:Pn} 
\end{eqnarray}  
Here, the coefficients $a^{(l)}_{nr}$ $(r=0,1,\cdots,n)$ satisfy the following
orthogonality relation 
\begin{eqnarray}\label{eq:orthogonality_conditionPn}
  \int\!\frac{d^3{\bf k} ~\omega^{(l)}_{\bf k}} {(2\pi)^3k_0} 
    P_{{\bf k} m}^{(l)} P_{{\bf k} n}^{(l)}=\delta_{mn}, 
  \label{eq:orthogonality_conditionPn_(a)}
\end{eqnarray} 
where $\omega^{(l)}_{\bf k}$ is an $l$-dependent weight factor defined by
\begin{eqnarray}
  \omega^{(l)}_{\bf k} &\equiv&
  \frac{{\cal W}^{(l)}}{(2l+1)!!} [\Delta^{\alpha\beta}k_{\alpha}k_{\beta}]^{l} f_{0{\bf k}},  
  \label{eq:definition_of_omega} 
\end{eqnarray}
and ${\cal W}^{(l)}$ in eq.(\ref{eq:definition_of_omega}) 
is a normalization factor (See Appendix \ref{app:E}).  
The orthogonal condition of eq. (\ref{eq:orthogonality_conditionPn_(a)}) gives a 
relation between the coefficients $a_{nr}^{(l)} $ and the integral as follows:
\begin{eqnarray}
   I^{(l)}_r \equiv \int \!\frac{d{\bf k}^3 \omega^{(l)}_{\bf k}}{(2\pi)^3k^0} (E_{\bf k})^r, 
\end{eqnarray}  
that is, 
\begin{eqnarray}
  \left( \begin{array}{llllll}
  1   & I^{(l)}_1 && \ldots &&  I^{(l)}_{n}    \\
  I^{(l)}_1 & I^{(l)}_2 && \ldots &&  I^{(l)}_{n+1} \\ 
  \hdotsfor{6} \\ 
 I^{(l)}_{n-1} & I^{(l)}_{n} && \ldots && I^{(l)}_{2n-1} \\ 
 I^{(l)}_n & I^{(l)}_{n+1} && \ldots && I^{(l)}_{2n} \\
\end{array}   \right) 
\left( \begin{array}{c} 
 a^{(l)}_{n0} \\  a^{(l)}_{n1} \\ \cdots \\  a^{(l)}_{n n-1} \\  
 a^{(l)}_{nn} \\ 
 \end{array} \right)  = 
\left( \begin{array}{c} 
 0 \\  0\\ \cdots \\  0 \\  
 1/a^{(l)}_{nn} \\ 
 \end{array} \right).
 \label{eq:orthogonal_matrix}
\end{eqnarray} 
The explicit form of $a_{nm}^{(l)}$ is given in Appendix \ref{app:G}. 
Now,  
$\lambda_{\bf k}^{\langle \mu_1\cdots \mu_l \rangle}$ 
and $\bar{\lambda}_{\bf k}^{\langle \mu_1\cdots \mu_l \rangle}$ 
can be 
expanded by using the polynomial function 
$P_{{\bf k}n}^{(l)}(E_{\bf k})$, as the following equation: 
\begin{subequations}
\begin{eqnarray}\label{eq:lambda_k}
    \lambda_{\bf k}^{\langle \mu_1\cdots \mu_l \rangle} &=&
     \sum_{n=0}^{\infty}
     c_{n}^{\langle \mu_1\cdots \mu_{l} \rangle}
     P_{{\bf k}n}^{(l)}, \label{eq:lambda_k_for_inf} 
     \\ 
 \bar\lambda_{\bf k}^{\langle \mu_1\cdots \mu_l \rangle} &=&
     \sum_{n=0}^{\infty}
     \bar c_{n}^{\langle \mu_1\cdots \mu_{l} \rangle}
     P_{{\bf k}n}^{(l)}.
\end{eqnarray}
\end{subequations}
Note that, without loss of generality, one can set $P_{{\bf k} 0}^{(l)} \equiv 1$ 
for arbitrary $l$, which is equivalent to 
$a^{(l)}_{00}=1$. 
For anti-particles, the normalization factor 
$\bar {\cal W}^{(l)}$ can be similarly defined (i.e., by 
replacement of $\mu\to -\mu$). 
The weight factor appeared in eq.(\ref{eq:definition_of_omega}) 
has no chemical potential dependence and we do not therefore need 
to introduce a factor $\bar{\omega}_{\bf k}^{(l)}$.
Substituting eq.(\ref{eq:lambda_k_for_inf}) into eq.(\ref{eq:expansion-lambda1}), 
multiplying a factor $f_{{\bf k}0} ~P_{{\bf k} m}^{(l')}~k^{\langle \mu_1}\cdots k^{\mu_{l'} \rangle}$ 
(with fixed $l'$) on the both side of it, and 
integrating over whole ${\bf k}$ space, we have 
\begin{eqnarray}
  \int\frac{d{\bf k}^3 \phi_{\bf k} f_{0{\bf k}}}{(2\pi)^3k^0} ~P_{{\bf k}n}^{(l')}~k^{\langle \nu_1}\cdots k^{\nu_{l'} \rangle}
  = \sum_{l=0}^{\infty}\sum_{n=0}^{\infty} 
  c_{n}^{\langle \nu_1\cdots \nu_{l} \rangle}
     P_{{\bf k}n}^{(l)} P_{{\bf k}n}^{(l')}~
     k^{\langle \nu_1}\cdots k^{\nu_{l'} \rangle} k_{\langle \mu_1}\cdots k_{\mu_l \rangle}. 
\end{eqnarray}
We then apply the orthogonal condition of eq.(\ref{eq:orthogonality_condition_final}) 
and (\ref{eq:orthogonality_conditionPn}) to the above 
expression, we obtain 
\begin{subequations}\label{eq:cn1} 
 \begin{eqnarray}
   c_{n}^{\langle \nu_1\cdots \nu_{l} \rangle} &=&
    \frac{{\cal W}^{(l)}}{l!}
    \int\frac{d{\bf k}^3 \phi_{\bf k} f_{0{\bf k}}}{(2\pi)^3k^0} ~ 
     P_{{\bf k}n}^{(l)}~k^{\langle \nu_1}\cdots k^{\nu_{l} \rangle}, \quad 
\end{eqnarray} and similarly for anti-particle part,   
\begin{eqnarray}
  \bar c_{n}^{\langle \nu_1\cdots \nu_{l} \rangle} &=&
    \frac{\bar{\cal W}^{(l)}}{l!}
    \int\frac{d{\bf k}^3 \bar\phi_{\bf k} \bar f_{0{\bf k} }}{(2\pi)^3k^0} ~ 
     P_{{\bf k}n}^{(l)}~k^{\langle \nu_1}\cdots k^{\nu_{l} \rangle}. \quad 
\end{eqnarray} 
\end{subequations} 
Substituting the definition of $P^{(l)}_{{\bf k}n}$ given by eq.(\ref{eq:Pn})  
into eq.(\ref{eq:cn1}) and denoting
\begin{subequations} \label{eq:rho_expression1}
\begin{eqnarray}
   \rho^{\mu_1\cdots\mu_l}_r 
&\equiv& \int\frac{d{\bf k}^3 \phi_{\bf k} f_{0{\bf k}}}{(2\pi)^3k^0} ~ 
   (E_{\bf k})^r  k^{\langle \nu_1}\cdots k^{\nu_{l} \rangle},\label{eq:rho1}\\
    \bar\rho^{\mu_1\cdots\mu_l}_r 
&\equiv& \int\frac{d{\bf k}^3 \bar\phi_{\bf k}\bar f_{0{\bf k}}}{(2\pi)^3k^0} ~ 
   (E_{\bf k})^r  k^{\langle \nu_1}\cdots k^{\nu_{l} \rangle},\label{eq:rho2}
\end{eqnarray}
\end{subequations}
 the following expressions are obtained (for $n=0,1,2,\cdots$)
\begin{subequations}\label{eq:cn2} 
\begin{eqnarray}
  c_n^{\langle \mu_1\cdots\mu_l\rangle} &=&
  \frac{{\cal W}^{(l)}}{l!}
  \sum_{m=0}^{n} a_{nm} ~\rho^{\mu_1\cdots\mu_l}_m, \label{eq:cn2a}\\  
  \bar c_n^{\langle \mu_1\cdots\mu_l\rangle} &=&
  \frac{\bar{\cal W}^{(l)}}{l!}
  \sum_{m=0}^{n} a_{nm} ~\bar\rho^{\mu_1\cdots\mu_l}_m~.\label{eq:cn2b}
\end{eqnarray}
\end{subequations}
The  $l$-rank coefficient
tensors $c_n^{\langle \mu_1\cdots\mu_l\rangle}$ and 
$\bar{c}_n^{\langle \mu_1\cdots\mu_l\rangle}$ are 
given by the linear combinations of $\rho^{\mu_1\cdots\mu_l}_m$ and 
$\bar{\rho}^{\mu_1\cdots\mu_l}_m$, respectively. 
Note here that 
$\rho_2 +\bar{\rho}_2 = u_{\mu}\delta T^{\mu\nu} u_{\nu} = \Lambda$,   
$\rho^{\mu}_1 + \bar{\rho}^{\mu}_1=  \Delta^{\mu}_{\alpha} 
\delta T^{\alpha\beta} u_{\beta} \equiv W^{\mu}$, 
$\rho^{\mu\nu}_0+\bar{\rho}^{\mu\nu}_0= 
\delta T^{\langle \mu\nu\rangle} \equiv \pi^{\mu\nu}$, 
$\rho_1 - \bar{\rho}_1 =u_{\mu}\delta N^{\mu} \equiv\delta n$ and 
$\rho^{\mu}_0- \bar{\rho}^{\mu}_0=N^{\langle \mu\rangle}  \equiv V^{\mu}$, 
where $W^{\mu}$, $\pi^{\mu\nu}$, and $V^{\mu}$ are, respectively, 
energy flow, shear tensor, and baryon number flow of the dissipative fluid. 
Therefore, if we truncate the expansion 
$\lambda_{\bf k}^{\langle \mu_1\cdots \mu_l \rangle}=0$ for $l>2$ and  
$c_{n}=\bar c_{n}=0$ for $n>2$ (with $l=0$),  
$c^{\langle\mu\rangle}_{n} =\bar c^{\langle\mu\rangle}_{n} =0$ for $n>1$ (with $l=1$), and 
$c^{\langle\mu\nu\rangle}_{n} =\bar c^{\langle\mu\nu\rangle}_{n} =0$ for $n>0$ (with $l=2$), 
linear combination of the rest non-zero coefficients 
$c_n^{\langle \mu_1\cdots\mu_l\rangle}$ and 
$\bar{c}_n^{\langle \mu_1\cdots\mu_l\rangle}$can be related 
to dissipative hydrodynamical quantities. 
We hereafter denote  
\begin{subequations}
\begin{eqnarray}
    \lambda_{\bf k}^{*\langle \mu_1\cdots \mu_l \rangle} &=&
     \sum_{n=0}^{N_l}
     c_{n}^{\langle \nu_1\cdots \nu_{l} \rangle}
     P_{{\bf k}n}^{(l)}, \label{eq:lambda_k_for_Nl} 
     \\ 
 \bar\lambda_{\bf k}^{*\langle \mu_1\cdots \mu_l \rangle} &=&
     \sum_{n=0}^{N_l}
     \bar c_{n}^{\langle \nu_1\cdots \nu_{l} \rangle}
     P_{{\bf k}n}^{(l)},\label{eq:lambda_k_bar_for_Nl} 
\end{eqnarray} 
\end{subequations}
such coefficient tensor with terminated at $N_2=0, N_1=1, N_2=2$,  known 
as the possible lowest truncation scheme \cite{PhysRevD.85.114047}.   
Although the definition of the eq.(\ref{eq:rho_expression1}) links the deviation 
$\phi_{\bf k}(x)$ and $\bar\phi_{\bf k}(x)$ to the local dissipative fluid dynamical quantities 
via both $\rho^{\mu_1\cdots\mu_l}_m$ and $\bar{\rho}^{\mu_1\cdots\mu_l}_m$, 
one can also give linkage between 
$\lambda_{\bf k}^{*\langle \mu_1\cdots\mu_l\rangle}$ ($\bar\lambda_{\bf k}^{*\langle \mu_1\cdots\mu_l\rangle}$)
and $\rho_r^{\mu_1\cdots\mu_l}$ ($\bar\rho_r^{\mu_1\cdots\mu_l}$) 
(See Appendix \ref{app:F} for the derivation of the below equations) as the following; 
\begin{subequations} \label{eq:from_micro_to_macro} 
\begin{eqnarray}
\rho_r^{\mu_1\cdots\mu_l}
&=& \frac{l!}{(2l+1)!!} \! 
   \int\!  \frac{d^3{\bf k}}{(2\pi)^3} 
   \lambda_{\bf k}^{*\langle \mu_1\cdots\mu_l\rangle}
   [\Delta^{\alpha\beta}k_{\alpha}k_{\beta}]^l 
   E_{\bf k}^r f_{{\bf k}0}, 
   \label{eq:from_micro_to_macro1} \\ 
\bar\rho_r^{\mu_1\cdots\mu_l}
&=&
   \frac{l!}{(2l+1)!!} \! 
   \int\!\! \frac{d^3{\bf k}}{(2\pi)^3} 
   \bar\lambda_{\bf k}^{*\langle \mu_1\cdots\mu_l\rangle}
   [\Delta^{\alpha\beta}k_{\alpha}k_{\beta}]^l 
   E_{\bf k}^r \bar f_{{\bf k}0}. 
   \label{eq:from_micro_to_macro2} 
\end{eqnarray} 
\end{subequations}
%
Finally, 
linear combinations of the l.h.s of eq.(\ref{eq:from_micro_to_macro1}) and 
eq.(\ref{eq:from_micro_to_macro2}) for $l=0$ gives 
\begin{subequations}\label{eq:final_expressions}
\begin{eqnarray}
  \Lambda &=& \int\!\!\frac{d^3{\bf k}~E_{\bf k}^2 }{(2\pi)^3k_0}  
   \Big[ \lambda_{\bf k}^* f_{{\bf k}0}(x) + 
   \bar\lambda_k^* \bar f_{{\bf k}0}(x) \Big], 
   \label{eq:Lambda_bulk} 
   \\
\delta n &=& \int\!\!\frac{d^3{\bf k}~E_{\bf k}}{(2\pi)^3k^0}  
   \Big[ \lambda_{\bf k}^* f_{{\bf k}0}(x) -  
   \bar\lambda_{\bf k}^* \bar f_{{\bf k}0}(x) \Big],
   \label{eq:delta_n}  
\end{eqnarray} 
and the bulk pressure is given by 
\begin{eqnarray}  
   \Pi &=& \frac{1}{3} \int\!\!\frac{d^3{\bf k}~k^2}{(2\pi)^3k^0}  
   \Big[ \lambda_k^* f_{{\bf k}0}(x) + 
   \bar\lambda_{\bf k}^* \bar f_{{\bf k}0}(x) \Big] .   
   \label{eq:PI_bulk} 
\end{eqnarray} 
For energy and net baryon number flow (for $l=1$),  we have   
\begin{eqnarray}
 W^{\mu} 
&=& -\frac{1}{3} \int\!\!\frac{d^3{\bf k}~E_{\bf k} k^2 }{(2\pi)^3k^0}  
   \Big[ \lambda_{\bf k}^{*\langle \mu \rangle} f_{{\bf k}0}(x) + 
   \bar\lambda_{\bf k}^{*\langle \mu \rangle} \bar f_{{\bf k}0}(x) \Big] \equiv  0,  \label{eq:W}\\
 V^{\mu} 
&=& -\frac{1}{3} \int\!\!\frac{d^3{\bf k}~k^2 }{(2\pi)^3k^0}  
   \Big[ \lambda_{\bf k}^{*\langle \mu \rangle} f_{{\bf k}0}(x) -  
   \bar\lambda_{\bf k}^{*\langle \mu \rangle} \bar f_{{\bf k}0}(x) \Big], \label{eq:V} 
\end{eqnarray}
and the shear viscosity (for $l=2$) is  
\begin{eqnarray}
 \pi^{\mu\nu} 
&=& \frac{2}{15} 
   \int\!\!\frac{d^3{\bf k}~k^4 }{(2\pi)^3k^0}  
   \Big[ \lambda_{\bf k}^{*\langle \mu\nu \rangle} f_{{\bf k}0}(x) +  
   \bar\lambda_{\bf k}^{*\langle \mu\nu \rangle}  \bar f_{{\bf k}0}(x) \Big], \label{eq:pi} 
\end{eqnarray}
\end{subequations}
where $k^2={\bf k}\cdot{\bf k}= -k_{\alpha}k_{\beta}\Delta^{\alpha\beta}$.
Note that the heat flow vector $W^{\mu}$ must vanish by definition 
of the Landau frame. 
(On the other hand, if the so called Eckart frame was employed, $V^{\mu}\equiv 0$.)  
As seen in eqs.(\ref{eq:W}) and (\ref{eq:V}), the tensor coefficients 
$\lambda_{\bf k}^{*\langle \mu \rangle}$ and $\bar\lambda_{\bf k}^{*\langle \mu \rangle}$ 
are restricted in the ${\bf k}$ dependence 
by the selection of the Lorentz frame for defining the rest frame of the fluid. 
Note that, eqs.(\ref{eq:Lambda_bulk}) and (\ref{eq:delta_n}) offer a key to understand the separation
temperature $T$ and the separation chemical potential $\mu$. 
Thus, we have
\begin{subequations}\label{eq:offer_key}
\begin{eqnarray}
  \varepsilon&=&\varepsilon(T,\mu,\lambda^*_{\bf k},\bar{\lambda}^*_{\bf k}) =
  \varepsilon_{\rm eq}(T,\mu)+\Lambda, \\
  n&=&n(T,\mu,\lambda^*_{\bf k},\bar{\lambda}^*_{\bf k}) =
  n_{\rm eq}(T,\mu)+\delta n. 
\end{eqnarray}
\end{subequations} 
Recall that $T$ and $\mu$ are functions of $\phi$ and $\bar{\phi}$ 
as shown by eq.(\ref{eq:phi_functional}). However, $T$ and $\mu$ 
can only be expressed by the $l=0$ component, i.e., 
$\lambda_{\bf k}^*$ and $\bar{\lambda}_{\bf k}^*$ as observed in eq.(\ref{eq:offer_key}). 
Hence the separation temperature $T$ and the separation chemical potential $\mu$ 
can be expressed as a function of 
$\lambda_{\bf k}^*(x)$ and $\bar{\lambda}_{\bf k}^*(x)$. 
\begin{subequations} 
\begin{eqnarray}
    T(x) &=& T[\lambda_{\bf k}^*(x),\bar{\lambda}_{\bf k}^*(x);T_0,\mu_0], \\ 
    \mu(x) &=& \mu[\lambda_{\bf k}^*(x),\bar{\lambda}_{\bf k}^*(x);T_0,\mu_0]. 
\end{eqnarray} 
\end{subequations} 

\subsection{Entropy current of the non-equilibrium state and thermodynamic stability}
When the microscopic phase-space distribution function $f_{\bf k}(x)$ is expressed 
by eq.(\ref{eq:phi_functional}), 
the local entropy current is divided into three parts as follows     
\begin{eqnarray}
  s^{\mu}(x) &=&-\int\!\!\frac{d^3{\bf k}~k^{\mu}}{(2\pi)^3k^0}
 \Big[
  [f_{\bf k}\ln f_{\bf k}-f_{\bf k}] + [\bar f_{\bf k}\ln \bar f_{\bf k}-\bar f_{\bf k}]
 \Big]  \nonumber \\ 
&=& s^{\mu}_{\rm eq}(x) +\delta s_1^{\mu}(x) +\delta s_2^{\mu}(x), 
  \label{eq;entropy_current} 
\end{eqnarray} 
where 
$s^{\mu}_{\rm eq}$ represents the equilibrium part given by 
the separation temperature $T$ and chemical potential $\mu$. 
The remaining two terms,    
$\delta s_1^{\mu}$ and $\delta s_2^{\mu}$, are 
 given by
\begin{subequations}
\begin{eqnarray}
   \delta s_1^{\mu} &\equiv&
   -\int\!\!\frac{d^3{\bf k} ~k^{\mu} }{(2\pi)^3k^0} 
   \Big[ 
    [\phi_{\bf k}\ln f_{{\bf k}0} ] ~f_{{\bf k}0} 
 + [\bar\phi_{\bf k}\ln \bar f_{{\bf k}0} ] ~\bar f_{{\bf k}0} \Big],\\ 
  \delta s_2^{\mu} &\equiv&
   -\int\!\!\frac{d^3{\bf k} ~k^{\mu} }{(2\pi)^3k^0} 
   \Big[ 
     [(1+\phi_{\bf k})\ln (1+\phi_{\bf k})-\phi_{\bf k} ] ~f_{{\bf k}0} 
  +  [(1+\bar\phi_{\bf k})\ln (1+\bar \phi_{\bf k}) -\bar\phi_{\bf k} ] 
   ~\bar f_{{\bf k}0} \Big],  \nonumber \\
\end{eqnarray}
\end{subequations}
respectively. 
Since the entropy density should be maximum and stable 
in the limit of equilibrium, it must not include any linear 
terms of scalar off-equilibrium quantities such as  
$\delta n$, $\Lambda$, and $\Pi$ \cite{PhysRevC.80.054906}; 
\begin{eqnarray}
  \frac{\partial (\delta s_1^{\mu}u_{\mu})}{\partial \delta n} \bigg|_{\delta n=0} = 
  \frac{\partial (\delta s_1^{\mu}u_{\mu})}{\partial \Lambda} \bigg|_{\Lambda=0} =
  \frac{\partial (\delta s_1^{\mu}u_{\mu})}{\partial \Pi} \bigg|_{\Pi=0}  \equiv 0. 
  \label{eq:stability_condition}   
\end{eqnarray}
Note that, the thermodynamic stability condition defined in
eq.(\ref{eq:stability_condition}) must be satisfied not only approximately but {\it
exactly}.
Therefore, one must add all terms for $\lambda_{\bf k}$ and $\bar{\lambda}_{\bf k}$ 
without termination in the energy polynomial function $P_{{\bf k}n}^{(0)}$ 
at a finite $n=N_0(=2)$.     
Thus the first order correction of the off-equilibrium entropy current should be 
given as the following: 
\begin{eqnarray} 
  \delta s_1^{\mu} &=&
        \bigg[ \int\!\!\frac{d^3{\bf k}}{(2\pi)^3k^0} 
        \Big\{ -\alpha E_{\bf k}(\lambda_{\bf k}f_{{\bf k}0}-\bar\lambda_{\bf k}\bar f_{{\bf k}0})\Big\} 
      +\int\!\!\frac{d^3{\bf k}}{(2\pi)^3k^0} 
       \Big\{ \beta E^2_{\bf k}(\lambda_{\bf k}f_{{\bf k}0}+\bar\lambda_{\bf k}\bar f_{{\bf k}0})\Big\} \bigg]u^{\mu} 
       \nonumber \\
     &-&\frac{1}{3} \int\!\!\frac{d^3{\bf k}~k^2}{(2\pi)^3k^0} 
        \Big\{ -\alpha E_{\bf k}(\lambda_{\bf k}^{\langle\mu\rangle} f_{{\bf k}0}
                                -\bar\lambda_{\bf k}^{\langle\mu\rangle}\bar f_{{\bf k}0})\Big\} 
         -\frac{1}{3} \int\!\!\frac{d^3{\bf k}~k^2}{(2\pi)^3k^0} 
        \Big\{ \beta E_{\bf k}^2(\lambda_{\bf k}^{\langle\mu\rangle} f_{{\bf k}0}
                                +\bar\lambda_{\bf k}^{\langle\mu\rangle}\bar f_{{\bf k}0})\Big\} \nonumber \\
        &\equiv& [-\alpha\delta n +\beta\Lambda +\xi]u^{\mu} 
        +[-\alpha V^{\mu} +\beta W^{\mu} +\xi^{\mu} ]. 
    \label{eq:delta_s1}
\end{eqnarray}
Here, $\alpha \equiv \mu/T$, $\beta\equiv 1/T$ and 
the terms $\xi$ and $\xi^{\mu}$ 
are residual terms which were ignored by the truncation 
of the polynomial $P^{(l)}_{{\bf k}n}$ at $n=N_l$ in $\lambda_{\bf k}^{*}$ and $\bar\lambda_{\bf k}^*$; 
\begin{eqnarray}
  \xi^{\mu_1\cdots \mu_l} &\equiv& -\int\!\!\frac{d^3{\bf k} E_{\bf k}}{(2\pi)^3k^0}
 \Big[ \alpha\delta\Phi^{\langle\mu_1\cdots \mu_l\rangle} 
  -\beta E_{\bf k} \delta\Psi^{\langle\mu_1\cdots \mu_l\rangle}  \Big] ,
\end{eqnarray}
where we denote 
\begin{subequations}
\begin{eqnarray}
&&
\delta\Phi^{\langle\mu_1\cdots \mu_l\rangle}_{\bf k}  \equiv  
  \delta\lambda_{\bf k}^{\langle\mu_1\cdots \mu_l\rangle}  f_{{\bf k}0}
 ~-~\delta\bar\lambda_{\bf k}^{\langle\mu_1\cdots \mu_l\rangle}  \bar f_{{\bf k}0}, \\ 
&&
\delta\Psi^{\langle\mu_1\cdots \mu_l\rangle}_{\bf k}  \equiv  
  \delta\lambda_{\bf k}^{\langle\mu_1\cdots \mu_l\rangle}  f_{{\bf k}0}
 ~+~\delta\bar\lambda_{\bf k}^{\langle\mu_1\cdots \mu_l\rangle}  \bar f_{{\bf k}0},
\end{eqnarray}
\end{subequations}
and
\begin{subequations}
\begin{eqnarray}\label{eq:delta_lambda_k} 
     \delta \lambda^{\langle \nu_1\cdots \nu_{l} \rangle}_{\bf k} &\equiv&
     \lambda^{\langle \nu_1\cdots \nu_{l} \rangle}_{\bf k} - \lambda^{*\langle \nu_1\cdots \nu_{l} \rangle}_{\bf k}=
     \sum_{n=N_l+1}^{\infty}
     c_{n}^{\langle \nu_1\cdots \nu_{l} \rangle}
     P_{{\bf k}n}^{(l)}, \\ 
     \delta \bar{\lambda}^{\langle \nu_1\cdots \nu_{l} \rangle}_{\bf k} &\equiv&
     \bar \lambda^{\langle \nu_1\cdots \nu_{l} \rangle}_{\bf k} - \bar\lambda^{*\langle \nu_1\cdots \nu_{l} \rangle}_{\bf k}=
     \sum_{n=N_l+1}^{\infty}
     \bar{c}_{n}^{\langle \nu_1\cdots \nu_{l} \rangle}
     P_{{\bf k}n}^{(l)}. 
\end{eqnarray}
\end{subequations}
In the usual formulation using the Landau matching conditions, 
the factor $[-\alpha\delta n +\beta\Lambda+\xi] u^{\mu} \equiv 0$ 
because $\delta n\equiv 0$, $\Lambda \equiv 0$, 
and $\xi=0$. Therefore, in this case, one may write  
\begin{eqnarray}
 \delta s^{\mu}_1 u_{\mu}=0, \label{eq:entropy_1st_vanish}
\end{eqnarray} 
which prevents the entropy current from 
the occurrence of instability \cite{PhysRevC.85.014906,OsadaEPJA48}
(See also Ref.\cite{PhysRevC.80.054906}). 
However, in the case of eq.(\ref{eq:extended-matching}), 
the condition of eq.(\ref{eq:entropy_1st_vanish}) is 
\begin{eqnarray}
   -\alpha \delta n + \beta \Lambda +\beta \chi \Pi \equiv 0,  
    \label{eq:stability}
\end{eqnarray} 
where $\chi\equiv \xi/(\beta\Pi)$ 
 for the off-equilibrium state to be thermodynamically stable. 
We require that the condition defined by eq.(\ref{eq:stability}) is 
satisfied for both particles and antiparticles because each of the 
subsystems should be independently stable; 
\begin{subequations}\label{eq_chi_expression}
\begin{eqnarray}
&& -\alpha \delta n_+ + \beta\Lambda_+ +\beta\chi_+\Pi_+ =0, 
    \label{eq:stability_p}\\
&& +\alpha \delta n_- + \beta\Lambda_- +\beta\chi_-\Pi_- =0.
    \label{eq:stability_n} 
\end{eqnarray}
Here suffixes $+$ and $-$ appearing 
in eq.(\ref{eq_chi_expression}) denote the 
contributions from particles ($\lambda_{\bf k}^* f_{{\bf k}0}$ 
in eq.(\ref{eq:final_expressions}))
and antiparticles ($\bar{\lambda}_{\bf k}^* \bar{f}_{{\bf k}0}$ in 
eq.(\ref{eq:final_expressions})). Then $\chi$ in eq.(\ref{eq:stability}) as 
\begin{eqnarray}
  \chi\equiv 
  \frac{\chi_+\Pi_++\chi_-\Pi_-}{\Pi_++\Pi_-} . 
  \label{eq:chi_final}
\end{eqnarray} 
\end{subequations}
Integration by parts for the first and the second terms in eq.(\ref{eq:delta_n}) yields  
\begin{subequations} \label{eq:off-equilibrium_EoS}
\begin{eqnarray}
  \delta n_+ &=&
 -\int\!\!\frac{d^3{\bf k}}{(2\pi)^3k^0}  \frac{E_{\bf k} k}{3} 
   \frac{\partial \lambda_{\bf k}^*}{\partial k} f_{{\bf k}0}  +\frac{\Pi_+}{T},
   \label{eq:deltan1}\\
\delta n_- &=&
  -\int\!\!\frac{d^3{\bf k}}{(2\pi)^3k^0} 
    \frac{E_{\bf k} k}{3} \frac{\partial \bar\lambda_{\bf k}^*}{\partial k} \bar f_{{\bf k}0} 
  +\frac{\Pi_-}{T}, \label{eq:deltan2} 
\end{eqnarray} 
\end{subequations} 
respectively. Note that, eqs.(\ref{eq:deltan1}) and (\ref{eq:deltan2}) can be regarded as 
`Equations of State' for the off-equilibrium condition that  correspond to  
$P_{{\rm eq}}= n_{\rm eq} T$ in the equilibrium state.  
By combining eq.(\ref{eq:off-equilibrium_EoS}), eq.(\ref{eq:stability}) and eq.(\ref{eq:final_expressions}),   
one obtain a differential equation for $\lambda_{\bf k}^*$ as
\begin{eqnarray}
&&
   \frac{\alpha E_{\bf k}}{3\beta}\Big( k\frac{d\lambda_{\bf k}^*}{dk} \Big) 
   +\Big(E_{\bf k}^2-\frac{[\alpha-\chi_+]}{3}k^2\Big) \lambda_{\bf k}^* =0~.
   \label{eq:differential_fq_for_lambda} 
\end{eqnarray} 
 The above differential equation can be solved by the following 
integration form 
\begin{eqnarray}
&&  \int \frac{d\lambda_{\bf k}^*}{\lambda_{\bf k}^*} = 
   \beta [1-\frac{\chi_+}{\alpha}]  \int\!\! \frac{~k~dk}{\sqrt{k^2+m^2}}  
   -\frac{3\beta}{\alpha} \int\!\! \frac{\sqrt{k^2+m^2}}{k} ~dk, 
\end{eqnarray}
thus we obtain  
\begin{subequations}\label{eq:solution_lambda} 
\begin{eqnarray}
   \lambda_{\bf k}^* &=& C_{\gamma} 
   \Big[ \frac{E_{\bf k}+m}{k}\Big]^{+\frac{3}{\alpha}m\beta}  
   \exp \Big[(1-\frac{[3+\chi_+]}{\alpha}) 
   \beta E_{\bf k}\Big]. ~\qquad  \label{eq:lambda}
\end{eqnarray}
For $\bar\lambda_k^*$, a similar expression can be obtained;  
\begin{eqnarray}
   \bar\lambda_{\bf k}^* &=& C_{\gamma} 
   \Big[ \frac{E_{\bf k}+m}{k}\Big]^{-\frac{3}{\alpha}m\beta}  
   \exp \Big[(1+\frac{[3+\chi_-]}{\alpha}) 
   \beta E_{\bf k}\Big]. ~\qquad \label{eq:lambda_bar} 
\end{eqnarray} 
\end{subequations}
The constant $C_{\gamma}$ in eq.(\ref{eq:solution_lambda}) is 
an arbitrary integration constant. Physically, it determines the absolute 
values of off-equilibrium quantities such as $\Pi$ 
\footnote{If one substitutes eqs.(\ref{eq:lambda}) and (\ref{eq:lambda_bar}) 
to eq.(\ref{eq:PI_bulk}), the bulk pressure $\Pi$ seems to be independent of temperature 
$T$ and it depends only on the chemical potential $\mu$ at a glance. 
However, $\Pi$ depends on $T$ because, 
through the thermodynamical stability conditions 
eqs.(\ref{eq:stability_p}) and (\ref{eq:stability_n}), 
$\chi_+$ and $\chi_-$ depend on $T$ 
(See also Fig.\ref{fig:4}, for the case of $\chi$). 
}. 
Therefore, we require that the value of $C_{\gamma}$ satisfies 
\begin{eqnarray}
 \Pi = \gamma ~P_{\rm eq}(\alpha,\beta), 
 \label{eq:def_gamma} 
\end{eqnarray}
where $\gamma$ indicates the strength of the
off-equilibrium state.
Hence, an off equilibrium state can be specified by $\alpha$, $\beta$, 
and $\gamma$. 
It is possible to write
 $\chi_{\pm}\equiv -3 \pm\delta \chi$ because 
an exchange $\alpha \leftrightarrow -\alpha$ 
in eq.(\ref{eq:solution_lambda}) gives 
$\lambda_{\bf k}^* \leftrightarrow \bar{\lambda}_{\bf k}^*$. 
Then, using eqs.(\ref{eq_chi_expression}),  
we can express $\chi$ 
in a more simple form
\begin{eqnarray}
  \chi = -3 +\frac{\delta\chi \delta\Pi}{\Pi}, 
  \label{eq:chi_arround3} 
\end{eqnarray}
where $\delta \Pi \equiv (\Pi_+ - \Pi_-)$. \\
%

\subsection{Coefficient tensors of the higher 
rank and link of the initial condition of fluid}\label{sec:2c}  
In the possible lowest scheme, the coefficient tensor of 
the first and second rank, 
$\lambda_{\bf k}^{*\langle \mu \rangle}$ and $\lambda_{\bf k}^{*\langle \mu\nu \rangle}$, 
are respectively terminated 
at $N_1=1$ and $N_2=0$ in eq.(\ref{eq:lambda_k_for_Nl}) and 
eq.(\ref{eq:lambda_k_bar_for_Nl}).  Hence, we can write 
\begin{subequations}
\label{eq:lambda_rank12} 
\begin{eqnarray}
 &&\lambda_{\bf k}^{*\langle \mu \rangle} = 
   c_0^{\langle \mu \rangle} 
+ c_1^{\langle \mu \rangle} P_{{\bf k}1}^{(1)},\quad  
   \bar\lambda_{\bf k}^{*\langle \mu \rangle} =
   \bar c_0^{\langle \mu \rangle} 
+ \bar c_1^{\langle \mu \rangle} P_{{\bf k}1}^{(1)},  \label{eq:lambda_rank1}\quad \\
  &&  \lambda_{\bf k}^{*\langle \mu\nu \rangle} = 
   c_0^{\langle \mu\nu \rangle} ,\quad 
   \bar \lambda_{\bf k}^{*\langle \mu\nu \rangle} = 
   \bar c_0^{\langle \mu\nu \rangle} ,\label{eq:lambda_rank2} \quad 
\end{eqnarray}
\end{subequations}
where $c_n^{\langle \mu_1\cdots\mu_l\rangle}$ and 
$\bar c_n^{\langle \mu_1\cdots\mu_l\rangle}$ are 
given by eq.(\ref{eq:cn2a}) and (\ref{eq:cn2b}) which are 
linear combination of $\rho^{\mu_1\cdots\mu_l}$ and 
$\bar \rho^{\mu_1\cdots\mu_l}$, 
\begin{subequations}\label{eq:c01}
\begin{eqnarray}
   &&  \frac{c_0^{\langle \mu\rangle}}{{\cal W}^{(1)}} = \rho^{\mu}_0,  \quad 
         \frac{\bar c_0^{\langle \mu\rangle}}{\bar{\cal W}^{(1)}}= \bar\rho^{\mu}_0, \label{eq:c0} \\
   &&  \frac{c_1^{\langle \mu\rangle}}{{\cal W}^{(1)}} = a_{10}^{(1)} \rho^{\mu}_0 +  a_{11}^{(1)} \rho^{\mu}_1,\quad 
   \frac{\bar c_1^{\langle \mu\rangle}}{\bar{\cal W}^{(1)}} =a_{10}^{(1)} \bar\rho^{\mu}_0 +  
   a_{11}^{(1)} \bar\rho^{\mu}_1, \label{eq:c1}
   \quad
\end{eqnarray} 
\end{subequations}
and the coefficient $a_{nm}^{(l)}$ are given by Appendix \ref{app:G}.
By substituting eq.(\ref{eq:lambda_rank1}),  
one obtains   
\begin{subequations}
\begin{eqnarray}
 && {\cal A}_0^1 ~\frac{c_0^{\langle \mu \rangle}}{{\cal W}^{(1)}}
  + \bar{\cal A}_0^1 ~\frac{\bar c_0^{\langle \mu \rangle}}{\bar{\cal W}^{(1)}}    
  + {\cal A}_1^1 ~\frac{c_1^{\langle \mu \rangle}}{{\cal W}^{(1)}} 
  + \bar{\cal A}_1^1 \bar ~\frac{c_1^{\langle \mu \rangle}}{\bar{\cal W}^{(1)}}   =0,\\
 && {\cal A}_0^0 ~\frac{c_0^{\langle \mu \rangle}}{{\cal W}^{(1)}} 
  - \bar{\cal A}_0^0 ~\frac{\bar c_0^{\langle \mu \rangle}}{\bar{\cal W}^{(1)}}    
  + {\cal A}_1^0 ~\frac{c_1^{\langle \mu \rangle}}{{\cal W}^{(1)}}  
  - \bar{\cal A}_1^0 ~\frac{\bar c_1^{\langle \mu \rangle}}{\bar{\cal W}^{(1)}}   =V^{\mu},
\end{eqnarray}
\end{subequations}
where (recall that $P_{{\bf k}0}^{(l)} \equiv 1$ and $a_{00}^{(l)}\equiv1$ 
for arbitrary $l$)
\begin{subequations}
\begin{eqnarray}
  {\cal A}_n^m &\equiv& 
-\frac{{\cal W}^{(1)}}{3} \int\!\!\frac{d^3{\bf k}k^2(E_{\bf k})^m}{(2\pi)^3k_0} 
  P_{{\bf k}n}^{(1)} f_{{\bf k}0}, \\
\bar{\cal A}_n^m &\equiv& 
-\frac{\bar{\cal W}^{(1)}}{3} \int\!\!\frac{d^3{\bf k}k^2(E_{\bf k})^m}{(2\pi)^3k_0} 
  P_{{\bf k}n}^{(1)} \bar f_{{\bf k}0}. 
\end{eqnarray}
\end{subequations} 
Since there are constraint conditions, 
i.e., $\rho^{\mu}_0-\bar\rho^{\mu}_0=V^{\mu}$ and 
$\rho^{\mu}_1+\bar\rho^{\mu}_1 (=W^{\mu})\equiv 0$, 
one can write
\begin{subequations}
\begin{eqnarray}
  \frac{\bar c_0^{\langle\mu\rangle}}{{\cal W}^{(1)}} &=&  
  \frac{c_0^{\langle\mu\rangle}}{{\cal W}^{(1)}} -V^{\mu}, \\
  \frac{\bar c_1^{\langle\mu\rangle}}{\bar{\cal W}^{(1)}} &=& 
-\frac{c_1^{\langle\mu\rangle}}{{\cal W}^{(1)}}
  +2a_{10}^{(1)} \frac{c_0^{\langle\mu\rangle}}{{\cal W}^{(1)}} -a_{10}^{(1)}V^{\mu}.  
\end{eqnarray}
\end{subequations}
Using the above equations to eliminate 
$\bar c_0^{\langle\mu\rangle}$ and $\bar c_1^{\langle\mu\rangle}$, one then obtains 
\begin{subequations}\label{eq:cn_coefficient_rank1}
\begin{eqnarray}
 c^{\langle\mu\rangle}_0 &=& {\cal W}^{(1)}~  
  \frac{
  \tilde{\cal A}^0_1 \tilde{\cal A}^1_v -\tilde{\cal A}^1_1(1-\tilde{\cal A}^0_v ) 
}{  \tilde{\cal A}^1_0 \tilde{\cal A}^0_1 -\tilde{\cal A}^1_1\tilde{\cal A}^0_0 }
~V^{\mu}, \\
 c^{\langle\mu\rangle}_1 &=& {\cal W}^{(1)}~  
  \frac{
  \tilde{\cal A}^1_0(1-\tilde{\cal A}^0_v ) -\tilde{\cal A}^0_0 \tilde{\cal A}^1_v
}{  \tilde{\cal A}^1_0 \tilde{\cal A}^0_1 -\tilde{\cal A}^1_1\tilde{\cal A}^0_0 }
~V^{\mu},  
\end{eqnarray} 
\end{subequations} 
where 
\begin{subequations}
\begin{eqnarray}
  \tilde {\cal A}^m_0 &\equiv& {\cal A}^m_0 +(-1)^{m+1} \bar {\cal A}^m_0 
 +(-1)^{m+1} 2a_{10}^{(1)} \bar{\cal A}^m_1,\quad  \\ 
  \tilde {\cal A}^m_1 &\equiv& {\cal A}^m_1 +(-1)^{m} \bar{\cal A}^m_1,  \\
  \tilde {\cal A}^m_v &\equiv& \bar{\cal A}^m_0 + a_{10}^{(1)} \bar{\cal A}^m_1. 
\end{eqnarray}
\end{subequations}
Similarly, substituting eq.(\ref{eq:lambda_rank2}) into eq.(\ref{eq:pi}), we also obtain  
\begin{eqnarray}
    {\cal D}_0^0 ~\frac{2! c_0^{\langle \mu\nu \rangle}}{{\cal W}^{(2)}}
  + \bar{\cal D}_0^0 ~\frac{2! \bar c_0^{\langle \mu\nu \rangle}}{\bar{\cal W}^{(2)}}     
  =\pi^{\mu\nu}, 
\end{eqnarray} 
where 
\begin{subequations}
\begin{eqnarray}
  {\cal D}_n^m &\equiv& 
  \frac{{\cal W}^{(2)}}{15} \int\!\!\frac{d^3{\bf k}k^4 (E_{\bf k})^m}{(2\pi)^3k_0} 
  P_{{\bf k}n}^{(2)} f_{{\bf k}0}, \\
\bar{\cal D}_n^m &\equiv& 
 \frac{\bar{\cal W}^{(2)}}{15} \int\!\!\frac{d^3{\bf k}k^4 (E_{\bf k})^m}{(2\pi)^3k_0} 
  P_{{\bf k}n}^{(2)} \bar f_{{\bf k}0}. 
\end{eqnarray}
\end{subequations} 
Because we have
\begin{eqnarray}
  \frac{2!c_0^{\langle\mu\nu\rangle}}{{\cal W}^{(2)}} 
+\frac{2!\bar c_0^{\langle\mu\nu\rangle}} {\bar{\cal W}^{(2)}} = 
  \rho^{\mu\nu}+ \bar\rho^{\mu\nu} =\pi^{\mu\nu},
\end{eqnarray} 
we obtain 
\begin{eqnarray}
       c^{\langle\mu\nu\rangle}_0 =
   \frac{1}{2!}{\cal W}^{(2)}
   \frac{1-\bar{\cal D}^0_0} {{\cal D}^0_0-\bar{\cal D}^0_0}~\pi^{\mu\nu}, \quad
\bar c^{\langle\mu\nu\rangle}_0 =
    \frac{1}{2!}\bar{\cal W}^{(2)}
   \frac{{\cal D}^0_0-1}{{\cal D}^0_0-\bar{\cal D}^0_0 }~\pi^{\mu\nu}.\quad 
\label{eq:cn_coefficient_rank2} 
\end{eqnarray}
As seen in eq.(\ref{eq:cn_coefficient_rank1}) and (\ref{eq:cn_coefficient_rank2}), 
one can connect the coefficient tensors 
$\lambda_{\bf k}^{*\langle\mu\rangle}$, 
$\bar\lambda_{\bf k}^{*\langle\mu\rangle}$, 
$\lambda_{\bf k}^{*\langle\mu\nu\rangle}$ and $\bar\lambda_{\bf k}^{*\langle\mu\nu\rangle}$ 
with $\delta N^{\langle\mu\rangle}$ and $\delta T^{\langle\mu\nu\rangle}$ 
as the following; 
\begin{subequations} \label{eq:lamda_tensor_rank12}
\begin{eqnarray}
 \lambda_{\bf k}^{*\langle\mu\rangle} &=&  
 \frac{1}{\tilde{\cal A}^1_0 \tilde{\cal A}^0_1 -\tilde{\cal A}^1_1\tilde{\cal A}^0_0 } 
 \Big[ (\tilde{\cal A}^0_1-\tilde{\cal A}^0_0 P^{(1)}_{{\bf k}1})\tilde{\cal A}^1_v 
 -(\tilde{\cal A}^1_1 -\tilde{\cal A}^1_0 P^{(1)}_{{\bf k}1})
   (1-\tilde{\cal A}^0_v ) \Big]
~\delta N^{\langle\mu\rangle}, \\
\bar\lambda_{\bf k}^{*\langle\mu\rangle} &=&  
 \frac{1}{\tilde{\cal B}^1_0 \tilde{\cal B}^0_1 -~\tilde{\cal B}^1_1\tilde{\cal B}^0_0 } 
\Big[ (\tilde{\cal B}^1_1 -\tilde{\cal B}^1_0 P^{(1)}_{{\bf k}1}) (1-\tilde{\cal B}^0_v )
 -~(\tilde{\cal B}^0_1-\tilde{\cal B}^0_0 P^{(1)}_{{\bf k}1})\tilde{\cal B}^1_v
\Big]
~\delta N^{\langle\mu\rangle}, \\
  \lambda_{\bf k}^{*\langle\mu\nu\rangle}&=&
   \frac{1-\bar{\cal D}^0_0} {{\cal D}^0_0-\bar{\cal D}^0_0}
~\delta T^{\langle\mu\nu\rangle}, \\ 
   \bar \lambda_{\bf k}^{*\langle\mu\nu\rangle} &=&
   \frac{{\cal D}^0_0-1}{{\cal D}^0_0-\bar{\cal D}^0_0 }
~\delta T^{\langle\mu\nu\rangle}, 
\end{eqnarray}
\end{subequations}
where ${\cal B}^m_n$ is obtained by exchange of ${\cal A}^m_n \leftrightarrow \bar{\cal
A}^m_n$; i.e., 
\begin{subequations}
\begin{eqnarray}
  \tilde {\cal B}^m_0 &\equiv& \bar{\cal A}^m_0 +(-1)^{m+1} {\cal A}^m_0 
 +(-1)^{m+1} 2a_{10}^{(1)} {\cal A}^m_1,\quad  \\ 
  \tilde {\cal B}^m_1 &\equiv& \bar{\cal A}^m_1 +(-1)^{m} {\cal A}^m_1,  \\
  \tilde {\cal B}^m_v &\equiv& {\cal A}^m_0 + a_{10}^{(1)} {\cal A}^m_1. 
\end{eqnarray}
\end{subequations}
It is shown that the coefficient tensor $\lambda^{*\langle\mu\rangle}_{\bf k}$ and  
$\lambda^{*\langle\mu\nu\rangle}_{\bf k}$ are given by the dissipative part of the 
baryon number 4-vector $N^{\mu}$ and 
energy-momentum tensor $T^{\mu\nu}$, respectively. 
Hence, if $N^{\mu}(x)$ and $T^{\mu\nu}(x)$ are given as an initial conditions, 
one can determine those coefficient tensors in the irreducible tensor expansion. 

Let us summarize the main points that we have been made in this section. 
For given energy-momentum tensor $T^{\mu\nu}$ and baryon number current 
vector $N^{\mu}$, one can find $\alpha\equiv \mu/T$, $\beta\equiv 1/T$ and $\gamma$ (the
strength of off-equilibrium) 
by solving the following equations simultaneously 
\begin{subequations}
\label{eq:formulation_summary} 
\begin{eqnarray}
  u_{\mu}T^{\mu\nu} u_{\nu} &=&\int\!\!\frac{d^3{\bf k}~E_{\bf k}^2 }{(2\pi)^3k_0}  
   \Big[ (1+\lambda_{\bf k}^*(\alpha,\beta;\gamma)) f_{{\bf k}0}(\alpha,\beta) + 
          (1+\bar\lambda_k^*(\alpha,\beta;\gamma)) \bar f_{{\bf k}0}(\alpha,\beta) \Big], 
           \label{eq:formulation_summary1} \qquad \\
  N^{\mu} u_{\mu} &=& \int\!\!\frac{d^3{\bf k}~E_{\bf k} }{(2\pi)^3k_0}  
   \Big[ (1+\lambda_{\bf k}^*(\alpha,\beta;\gamma)) f_{{\bf k}0}(\alpha,\beta) -  
          (1+\bar\lambda_k^*(\alpha,\beta;\gamma)) \bar f_{{\bf k}0}(\alpha,\beta) \Big], 
           \label{eq:formulation_summary2} \qquad \\
  -\frac{1}{3} \Delta_{\mu\nu} T^{\mu\nu} &=&
  \frac{1}{3} \int\!\!\frac{d^3{\bf k}~k^2 }{(2\pi)^3k_0}  
   (1+\gamma) \Big[ f_{{\bf k}0}(\alpha,\beta) +\bar f_{{\bf k}0}(\alpha,\beta) \Big], 
   \label{eq:formulation_summary3} 
\end{eqnarray}
\end{subequations} 
where $\lambda_{\bf k}^*$ and $\bar\lambda_{\bf k}^*$ are given by (\ref{eq:lambda}) 
and (\ref{eq:lambda_bar}), respectively. As seen in the eq.(\ref{eq:formulation_summary}), 
the temperature $T$ and chemical potential $\mu$ are connected with  
value of $u_{\mu}T^{\mu\nu}u_{\nu}$ and $u_{\mu}N^{\mu}$ via 
$\lambda_{\bf k}^*$ and $\bar \lambda_{\bf k}^*$ and they are also related with 
$-\Delta_{\mu\nu}T^{\mu\nu}/3$. 
Note that, on the other hand, 
the Landau matching condition directly connect $T$ and $\mu$ with those values 
regardless of the value of the bulk pressure $\Pi=-\Delta_{\mu\nu}T^{\mu\nu}/3$. 
If the above a set of equations 
eq.(\ref{eq:formulation_summary}) are solved and 
the separation temperature $T$ and chemical potential $\mu$ obtained, 
coefficients tensor of first and second rank are also obtained by using 
eq.(\ref{eq:lamda_tensor_rank12}). 
Unlike the case of the coefficient tensor of zero-rank, 
we found that those of the first and second rank 
are proportional to $\delta N^{\langle\mu\rangle}$ and 
$\delta T^{\langle\mu\nu\rangle}$, respectively. 

Before closing this section, comments concerning 
independent degrees of freedom of relativistic 
dissipative fluid with extended matching condition  
may be in order here.
The energy-momentum tensor $T^{\mu\nu}$ has 10 and 
the net baryon current $N^{\mu}$ has 4 independent 
degree of freedom.   
These quantities can be obtained from some kind of 
quark-gluon distribution function in 
early stage of relativistic heavy-ion collisions.  
For example, one may consider 
gluon distribution function for a glasma state \cite{McLerran:2014apa,Praszalowicz:2013fsa}. 
Once such a quark-gluon distribution functions ($f_{\bf k}$ and $\bar f_{\bf k}$) 
are introduced, those 14 degrees of freedom 
\footnote{These 14 degrees of freedoms are accounted for by $\varepsilon$, $n$, $u^{\mu}$ that satisfy 
$u^{\mu}u_{\mu}\equiv1$ (3 independent degree of freedom), 
the bulk pressure $\Pi$, and the dissipative tensor $\pi^{\mu\nu}$ 
and current $V^{\mu}$ that satisfy 
$\pi^{\mu\nu}u_{\mu}=0, \pi^{\mu}_{\mu}=0$ (5 independent degree of freedom) 
and $V^{\mu}u_{\mu}=0$ (3 independent degree of freedom).} 
are fixed by eq.(\ref{eq:energy-momentum_tensor}) and (\ref{eq:particle_current}). 
For the energy-momentum tensor $T^{\mu\nu}$ obtained, 
one extracts $\varepsilon$, $u^{\mu}$ from eq.(\ref{eq:Landau_frame}), 
$n$ from eq.(\ref{eq:current_decomposition}) 
and $P_{\rm eq}+\Pi$ from its definition $-\Delta_{\mu\nu}T^{\mu\nu}/3$ 
\footnote{The Landau matching condition separates $P_{\rm eq}$ and $\Pi$ 
by using an equation of state $P_{\rm eq}(\varepsilon,n)$ 
because $\varepsilon=\varepsilon_{\rm eq}$ and $n=n_{\rm eq}$ are assumed.}. 
Hence, 6 degrees of freedom are fixed.   
The other 8 degrees are fixed as follows: i.e., 
3 degrees of freedom for $V^{\mu}$ and 5 degrees for $\pi^{\mu\nu}$ 
are attribute to those for 
$\lambda_{\bf k}^{*\langle\mu\rangle}$ and $\lambda_{\bf k}^{*\langle \mu\nu\rangle}$, 
respectively, in the irreducible tensor expansion of the initial distribution function. 
See also Table \ref{tab:1}. 

The zero-th order coefficients $\lambda_{\bf k}^*$ and $\bar\lambda_{\bf k}^*$  
have been ignored so far because of the Landau matching condition, i.e., 
$\varepsilon=\varepsilon_{\rm eq}(\alpha,\beta)$ and $n=n_{\rm eq}(\alpha,\beta)$.  
If the condition is extended, $\lambda_{\bf k}^*$  and $\bar\lambda_{\bf k}^*$ not only 
lead new `internal degrees of freedom' (i.e., $\Lambda$ and $\delta n$) 
to the energy density $\varepsilon$ and the net baryon density $n$, 
but also link those $\Lambda$ and $\delta n$ with the bulk pressure $\Pi$.
Note that, by the extension of the matching condition, only one degree of 
freedom $\lambda_{\bf k}^*$ is newly introduced. 
(See eqs. (\ref{eq:Lambda_bulk})-(\ref{eq:PI_bulk}).   
Since $\bar\lambda_{\bf k}^*$ must be linked to $\lambda_{\bf k}^*$ 
by replacement of $\alpha \to -\alpha$, they are not independent degree of freedom in fact.)
The momentum ${\bf k}$ dependence for $\lambda_{\bf k}^*$ is actually 
determined 
by the thermodynamical stability condition eq.(\ref{eq:stability}) 
and its absolute values ($C_{\gamma}$ in the eq.(\ref{eq:solution_lambda})) 
can be determined by the bulk pressure $\Pi$, which is characterized by the parameter $\gamma$.
Through the constraint of eq.(\ref{eq:stability}), 
$\chi$ links to $\lambda_{\bf k}^*$ and it appears in the 
expression of eq.(\ref{eq:solution_lambda}). 
Hence, the parameter $\chi$ and $\lambda_{\bf k}^*$ are 
function of the $\alpha$, $\beta$ and $\gamma$.  
Although the extension of the matching condition leads a new degree of freedom 
$\lambda_{\bf k}^*$, it is able to be given by a function of other degrees of freedom 
(i.e., $\alpha, \beta$ and $\gamma$) due to the constraint of eq.(\ref{eq:stability}) 
with using eq.(\ref{eq:off-equilibrium_EoS}).  
Therefore, even for the case that the extended matching condition is used 
in the relativistic dissipative fluid model, 14 degrees of freedom are required 
which is the same as the number used in the case of the usual Landau matching condition. 
\begin{center}
\begin{table}
\newcommand{\lw}[1]{\smash{\lower2.0ex\hbox{#1}}}
\newcommand{\lww}[1]{\smash{\lower1.25ex\hbox{#1}}}
\renewcommand{\arraystretch}{1.25}
\caption{\label{tab:1} 
Separation of equilibrium part $\varepsilon_{\rm eq}$, $n_{\rm eq}$ (or $\alpha$, $\beta$) from an 
off-equilibrium state characterized by $T^{\mu\nu}$ and $N^{\mu}$ with using a scalar dissipative strength $\gamma$ 
in the Landau-frame.} 
\begin{tabular}{|c|c|c|c|}\hline
\lww{\footnotesize energy-momentum tensor}  &  
\lw{equation(s)} & \lww{\footnotesize kinetic parameter obtained} & 
\lww{\footnotesize thermodynamical} \\ 
{\footnotesize and net baryon current} & 
&{\footnotesize (distribution. function.)} &{\footnotesize and fluid quantity} \\ 
\hline \hline 
$\varepsilon=u_{\mu}T^{\mu\nu}u_{\nu}$ &&$\alpha$&
\lw{$\varepsilon_{\rm eq}, n_{\rm eq}$ } \\ \cline{1-1}\cline{3-3}
$n=N^{\mu}u_{\mu}$&\lw{eq.(61)}&$\beta$& \\ \cline{1-1}\cline{3-4}
\lw{$-\frac{1}{3}\Delta_{\mu\nu}T^{\mu\nu}\equiv (1+\gamma)P_{\rm eq}$} 
&&$\gamma$&\lw{$\Lambda,\delta n, \Pi$} \\ \cline{3-3} 
&&$\lambda_{\bf k}^*(\alpha,\beta;\gamma)$&\\ \hline 
$T^{\mu\nu} u^{\nu}=\varepsilon u^{\mu}$ &eq.(9a) &$u^{\mu}$& $u^{\mu}$ \\ \hline
$\Delta^{\mu}_{\lambda}T^{\lambda}_{\sigma}u^{\sigma}\equiv 0$
& ~{\footnotesize Landau frame}~& --- & $W^{\mu}\equiv 0$ \\ \hline
$N^{\langle\mu\rangle}$&~{\footnotesize eq.(59a) and (59b)}~&
$\lambda_{\bf k}^{*\langle\mu\rangle}$ & $V^{\mu}$ \\ \hline
$T^{\langle\mu\nu\rangle}$&~{\footnotesize eq.(59c) and (59d)}~
&$\lambda_{\bf k}^{*\langle\mu\nu\rangle}$ & $\pi^{\mu\nu}$\\ \hline
\end{tabular} 
\end{table}
\end{center}
\section{Numerical results and discussion}\label{sec:3}
In this section, we demonstrate the separation of the corresponding 
equilibrium energy density $\varepsilon_{\rm eq}(T,\mu)$ and 
net baryon density  $n_{\rm eq}(T,\mu)$ from 
those in an off-equilibrium state ($\varepsilon$ and $n$) 
provided that 
the strength of the off-equilibrium state $\gamma$ is fixed \footnote{
In this case, $\alpha$ and $\beta$ are obtained by solving eqs.(\ref{eq:formulation_summary1}) 
and (\ref{eq:formulation_summary2}) simultaneously with using fixed $\gamma$. 
The bulk pressure is then obtained by $\Pi=\gamma P_{\rm eq}(\alpha,\beta)$. 
Therefore, the demonstration presented here corresponding to finding the 
separation temperature $T$ and chemical potential $\mu$  
for fixed $u_{\mu}T^{\mu\nu}u_{\nu}$ and $N^{\mu}u_{\mu}$ 
with varying $-\frac{1}{3} \Delta_{\mu\nu} T^{\mu\nu}$.}.  
The boundary conditions (for $\gamma\to 0$ limit, i.e., $\Lambda=\delta n=0$)  
of the separation temperature and chemical potential, 
(which correspond to equilibrium temperature and chemical potential, 
respectively, because of $\gamma=0$)  
are determined by the Landau matching condition:
\begin{eqnarray}
    \varepsilon=\varepsilon_{\rm eq}(\alpha_0,\beta_0), \quad 
    n=n_{\rm eq}(\alpha_0,\beta_0), 
    \label{eq:boundary_condition2} 
\end{eqnarray}
where $\alpha_0\equiv \mu_0/T_0$, $\beta_0\equiv 1/T_0$. 
For all numerical results shown, 
the classical particle mass $m$ is 5 MeV.  

The extended matching conditions now reads   
\begin{subequations}\label{eq:lambda-deltan_2}
\begin{eqnarray}  
   \varepsilon-\varepsilon_{\rm eq}(\alpha,\beta)
&=&\Lambda(\alpha,\beta,\gamma),  \label{eq:simultanous_eq1}\\
   n-n_{\rm eq}(\alpha,\beta)
&=& \delta n(\alpha,\beta,\gamma). \label{eq:simultanous_eq2}
\end{eqnarray} 
\end{subequations}
Therefore, determination of the separation energy density and net baryon density is 
equivalent to solving the nonlinear simultaneous equations concerning 
$\alpha$ and $\beta$ for given $\varepsilon$, $n$, and with fixing $\gamma$.    
We require that total energy and 
net baryon number in the fluid cell should be 
independent of $\gamma$; i.e., 
although $\gamma$ changes form of the distribution function, 
the total energy and net baryon number in the local fluid cell are 
set to be unchanged.   
\begin{figure}
\begin{center}
\includegraphics[width=14cm]{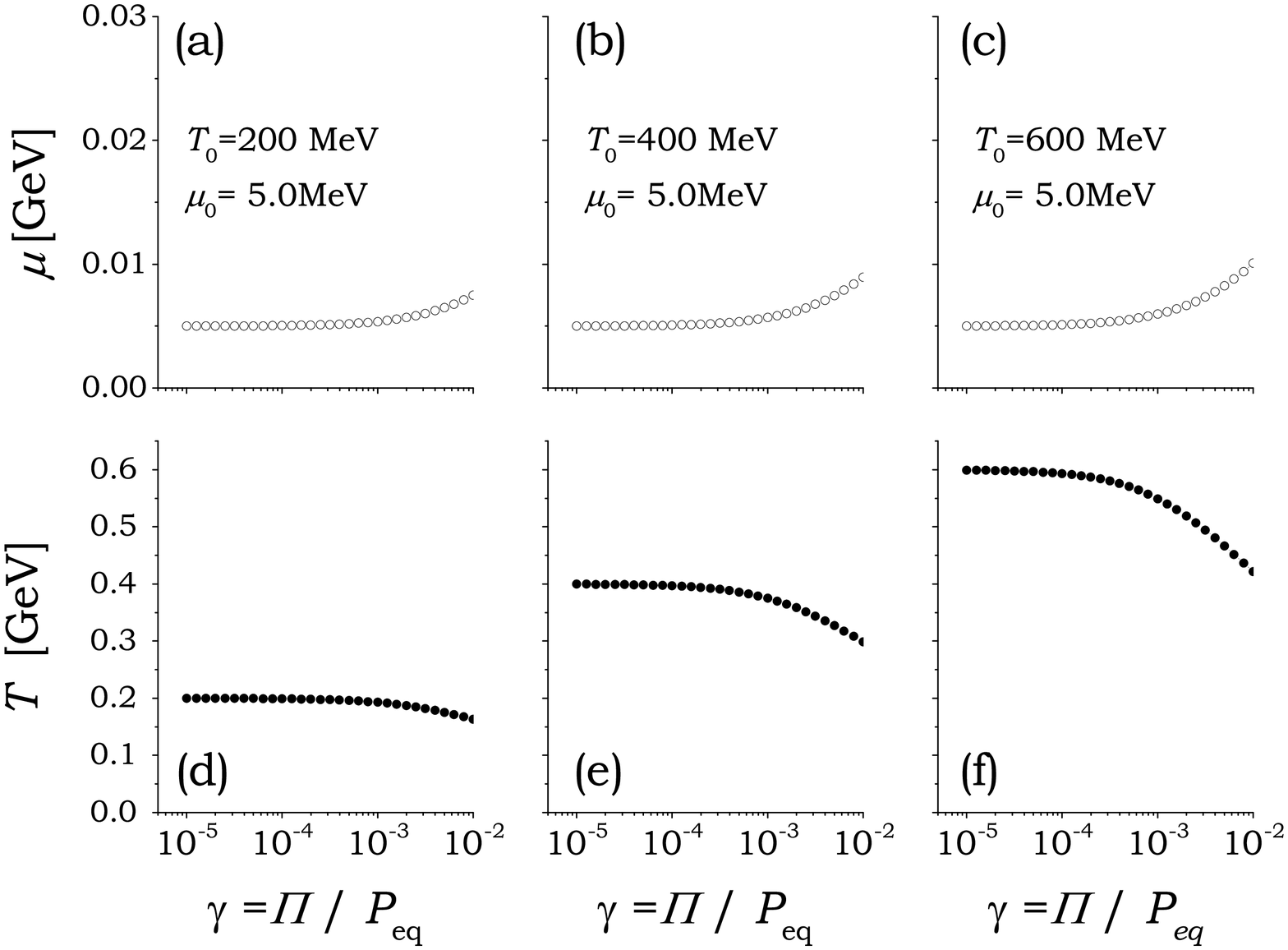}
\caption{\label{fig:1} 
The separation chemical potential $\mu$ [GeV] (upper panels (a), (b), and (c))
and separation temperature $T$ [GeV] (lower panels (d), (e) and (f)) for 
an off-equilibrium fluid as functions of $\gamma=\Pi/P_{\rm eq}$.
The $T_0$=0.2 GeV for figure (a) and (d), $T_0$=0.4 GeV, for (b) and (e), and  
$T_0$=0.6 GeV for (c) and (f), while the $\mu_0$=5.0 MeV is fixed for all panels.}
\includegraphics[width=14cm]{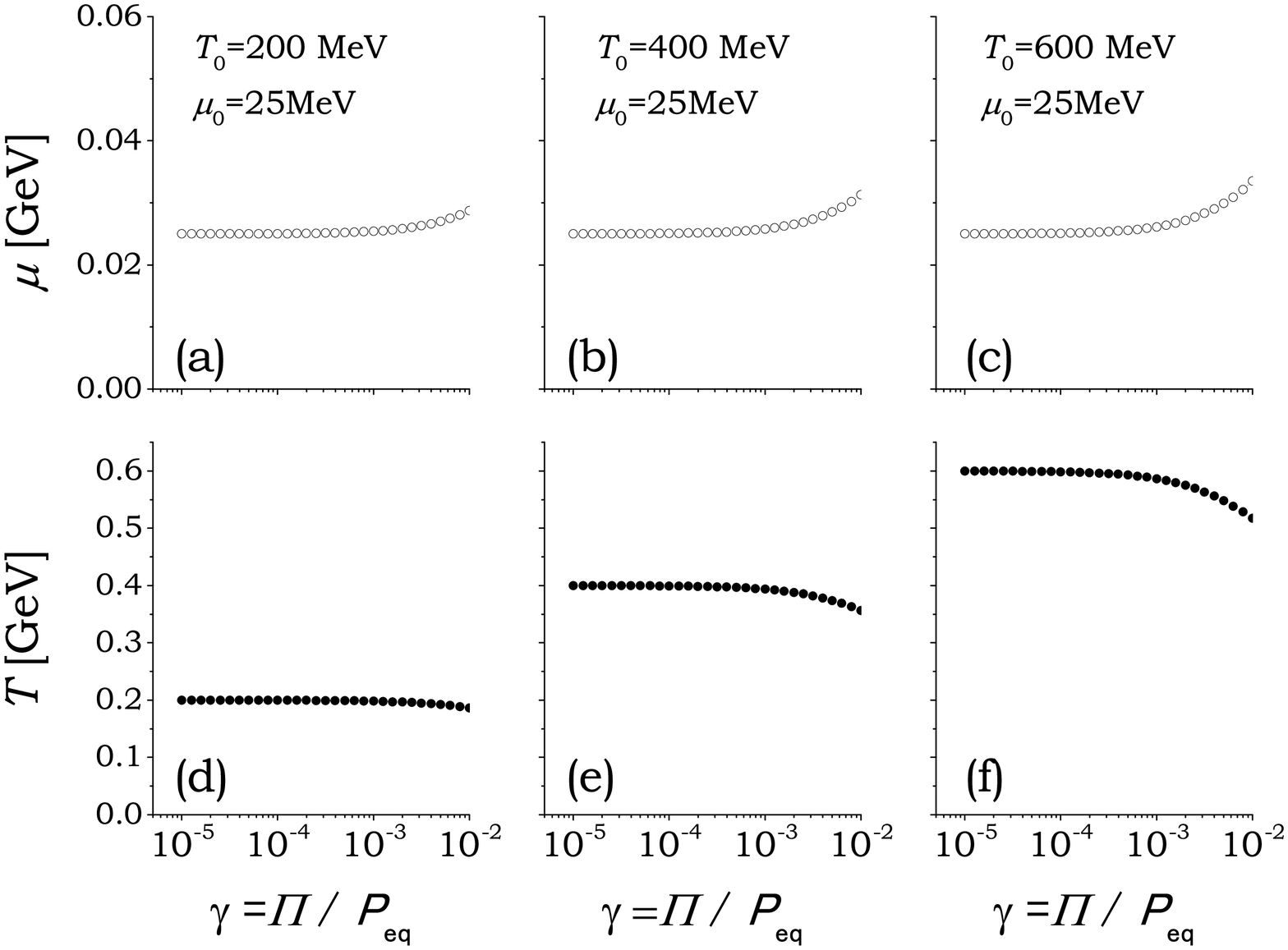}
\caption{\label{fig:2} The same as Fig.1 but $\mu_0$=25 MeV.}
\end{center}
\end{figure}
Figures \ref{fig:1} and \ref{fig:2} show the separation temperature $T$ 
and separation chemical potential $\mu$ as a function of 
$\gamma=\Pi/P_{\rm eq}$, 
$\mu_0$=5.0 MeV (Fig.\ref{fig:1}) and $\mu_0$=25 MeV (Fig.\ref{fig:2}). 
$T_0$ is fixed at 200 MeV [panels (a) and (d) of Figs.\ref{fig:1} and \ref{fig:2}], 
400 MeV  [panels (b) and (e) of Figs.\ref{fig:1} and \ref{fig:2}], and 
600 MeV  [panels (c) and (f) of Figs.\ref{fig:1} and \ref{fig:2}]. 
For very small $\gamma$ region ($\gamma \lesssim 10^{-4}$) 
in both Figs. \ref{fig:1} and \ref{fig:2}, one can observe 
that  $T$ and $\mu$ are almost constant with $T\approx T_0$ and $\mu\approx\mu_0$. 
In this region, the Landau matching condition works well. 
However, in the region $\gamma \gtrsim 10^{-3}$, 
the separation temperature $T$ decreases with increasing $\gamma$ and 
the separation chemical potential increases. 
The reason for the decrease of the separation temperature $T$ is 
that the total energy of the fluid is partially used in non-thermal motion 
such as tiny scale turbulent flow as discussed in Section \ref{sec:1}. 
On the other hand, the separation chemical potential 
increases as $\gamma$ increases for an off-equilibrium system. 
This is because 
of the constraint of conservation of the net-baryon number, 
i.e., the difference between the number of particles $n_+$ 
and anti-particles $n_-$ must be fixed while the total number of 
particles and anti-particles $n_+ + n_-$ decreases due to the decrease in
temperature $T$.  
\begin{figure}
\begin{center} 
\includegraphics[width=11cm]{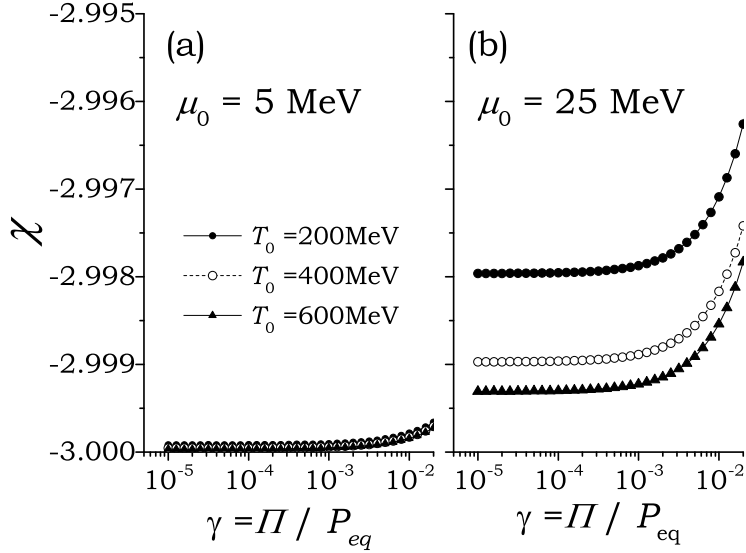}
\caption{\label{fig:3} The $\gamma$ dependence of the factor 
$\chi$ eq.(\ref{eq:chi_final}) with the equilibrium temperature 
$T_0$=200, 400, and 600 MeV, and the equilibrium chemical potential 
$\mu_0$=5 MeV (panel (a)) and for $\mu_0$=25 MeV (panel (b)). 
In the limit of $\gamma\to 0$ and $\mu_0\to 0$, the $\chi$ 
approaches to $-3$.}
\end{center} 
\end{figure}
Figure \ref{fig:3} shows the $\gamma$ dependence of the $\chi$ 
parameter for different $\mu_0$ and $T_0$. 
In the limit $\mu_0\to0$ $\chi$ approaches $-3$ 
because $\delta \chi \to 0$. 
We have already reported the result $\Lambda=3\Pi$ for 
the baryon free case in the NeXDC correspondence \cite{PhysRevC.77.044903}, 
as well as the hydrodynamical model in the presence of a long-range 
correlation \cite{PhysRevC.81.024907}. 
When the strength of the off-equilibrium state 
$\gamma\equiv \Pi/P_{\rm eq}$ is larger than 
around $\sim 1/100$, the value of $\chi$ increases rapidly. 
This is because the bulk pressure gap between $\Pi_+$ 
and $\Pi_-$ becomes large in the region of $\gamma > 1/100$. 
Recall that the parameter $\chi$ is derived from $\xi$ ($\xi=\chi\beta\Pi$), 
which is 
the residual term neglected by truncation 
of the energy polynomial function $P_{{\bf k}n}^{(l)}$.
This means that 
not only quadratic, but also higher order energy 
dependences of both $\lambda_{\bf k}$ and $\bar{\lambda}_{\bf k}$ 
play an important role in the thermodynamical stability of the 
off-equilibrium entropy current. 
These higher order contributions 
in the energy polynomial function may result in the existence of the 
tiny turbulent flow in the fluid. 
For the finite net baryon number case, we have also obtained 
the same expression as given in eq.(\ref{eq:stability}) 
\cite{PhysRevC.85.014906,OsadaEPJA48}. We found there that the value of $\chi$ plays 
an important role to restore not only the thermodynamical stability but also 
the causality of the solution obtained from the relativistic dissipative fluid 
dynamical equations. 
In ref.\cite{PhysRevC.85.014906,OsadaEPJA48}, the value of $\chi$ is introduced 
phenomenologically into the expression for an off-equilibrium entropy current, 
however, in this article, we illustrate the result numerically using kinetic theory 
 as shown in Fig.\ref{fig:3}.

\begin{figure} 
\begin{center}
\includegraphics[width=16cm]{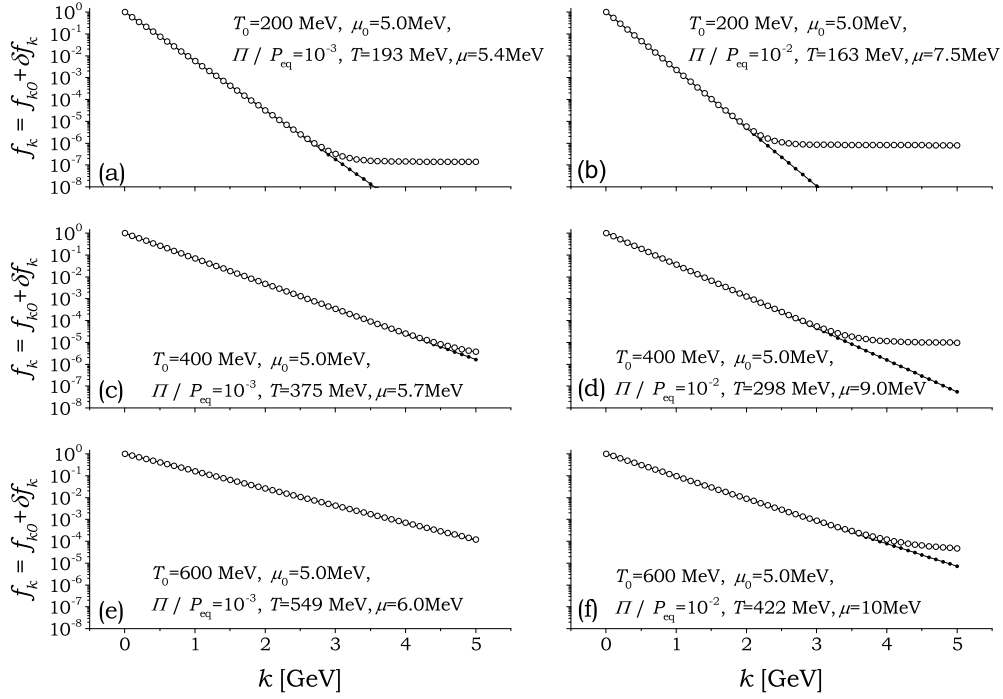}
\caption{\label{fig:4}
The off-equilibrium distribution function $f_{{\bf k}0}+\delta f_{\bf k} =
(1+\lambda_k^*)f_{{\bf k}0}$ 
for $\mu_0$= 5 MeV, 
$T_0$= 200 MeV (panel (a)), 400 MeV (panel (c)), 
and 600 MeV (panel (e)) in the case of $\gamma=10^{-3}$.
The right side panels (b), (d) and (f) are the same as 
(a), (c) and (e), respectively, but for $\gamma=10^{-2}$.   
The straight line (with closed circle) 
is the Boltzmann distribution $f_{{\bf k}0}$ with 
the separation temperature $T$, and the separation chemical potential 
$\mu$ are shown in each panel. The open circles shows the off-equilibrium 
distribution function $f_{{\bf k}0}+\delta f_{\bf k}$.} 
\end{center}
\end{figure}

In figures \ref{fig:4}, it is shown that the first order (scalar) correction $\lambda_{\bf k}^*$ 
to the distribution function 
for $\gamma = 10^{-3}$ and $\gamma=10^{-2}$
with $T_0$ =200, 400, and 600 MeV and $\mu_0$=5.0 MeV.  
As seen in Figs.\ref{fig:4}, 
the off-equilibrium correction increases more rapidly in the large momentum region. 
This tendency is seen more explicitly as $\gamma$ increases and as 
the separation temperature $T$ decreases. 
In addition to the correction $\lambda_{\bf k}^*$, 
one need to take vector and tensor contribution, i.e., 
$\lambda_{\bf k}^{*\langle\mu\rangle}k_{\langle\mu\rangle}$ and 
$\lambda_{\bf k}^{*\langle\mu\nu\rangle}k_{\langle\mu}k_{\nu\rangle}$, respectively,  
into account to estimate the correction for the distribution function. 
However, to do this, one need set full energy-momentum tensor and net-charge 
current vector and this is outside the scope of the present manuscript. 
We plan to discuss this subject elsewhere. 
\section{Summary and concluding remarks}\label{sec:4} 
We have proposed a novel way to specify the initial conditions for a
dissipative fluid dynamical model as an alternative to the so-called Landau matching 
condition eq.(\ref{eq:normal-matching1}), employed in most of the literature published so
far. 
The large expansion rate of matter produced in ultra-relativistic heavy-ion 
collisions may prevent the local equilibrium condition eq.(\ref{eq:perfect_condition})
from holding true.  Specifically, the microscopic collision time scale is 
not much shorter than any macroscopic evolution time scale 
because the macroscopic evolution time scale  is considerably 
reduced due to the large gradients in thermodynamical 
quantities when compared to a perfect fluid. 
In cases characterized by eq.(\ref{eq:inparfect_condtion}), 
the distribution functions are defined by eq.(\ref{eq:modification_in_distribution}) 
\begin{eqnarray*}
   f_{\bf k}(x)= f_{{\bf k}0}[T(x),\mu(x)] + \delta f_{\bf k}(x), \\ 
   \bar{f}_{\bf k}(x)= \bar{f}_{{\bf k}0}[T(x),\mu(x)] + \delta \bar{f}_{\bf k}(x). 
\end{eqnarray*}
For a given off-equilibrium state specified by the energy-momentum 
tensor given in eq.(\ref{eq:energy-momentum_tensor})
and net baryon number current defined according to eq.(\ref{eq:particle_current}), 
the Landau matching procedure equals 
the energy density and net baryon number density of the 
off-equilibrium state with those in an equilibrium of  the form 
$T_{\rm eq}^{\mu\nu}$ and $N_{\rm eq}^{\mu}$,   
\begin{eqnarray*}
   u_{\mu}T^{\mu\nu}u_{\nu} = u_{\mu}T^{\mu\nu}_{\rm eq} u_{\nu}, \quad 
   N^{\mu}u_{\mu} = N^{\mu}_{\rm eq} u_{\mu}, 
\end{eqnarray*}
which imposes strict restrictions on the distributions $\delta f_{\bf k}(x)$ and $\delta
\bar{f}_{\bf k}(x)$. 
Although the physical meaning of the matching (fitting) condition is unclear,  
it provides the temperature $T$ and chemical potential $\mu$ 
for a given off-equilibrium energy density $\varepsilon=\varepsilon_{\rm eq}(T,\mu)$ 
and net baryon density $n=n_{\rm eq}(T,\mu)$. 
One may regard the Landau matching condition as {\it just a definition} 
for specifying an off-equilibrium state.
However, it should be noted here that other off-equilibrium 
quantities such as the bulk pressure $\Pi$ are undoubtedly 
distorted by the condition. 
Therefore, we assert that the Landau matching condition be 
 relaxed as shown in eq.(\ref{eq:extended-matching}), 
\begin{eqnarray*}
     \delta T^{\mu\nu} u_{\mu} u_{\nu} 
     =\Lambda\ne 0, \quad \delta n^{\mu} u_{\mu} =\delta n \ne 0. 
\end{eqnarray*}  
Then, $\Lambda$ and $\delta n$ obtain their entities of physical existence, 
i.e., they are the energy density and baryon number density of 
the tiny scale (less than cell size scale) 
disturbance flow caused by the rapid expansion of the viscous fluid.  

An important consequence of the irreducible tensor expansion 
for the off-equilibrium distribution function is that 
 scalar quantities such as $\Lambda$, $\delta n$, and $\Pi$ can 
be written as functions of the separation temperature $T$, 
chemical potential $\mu$, and the 0-th order scalar expansion coefficients 
$\lambda_{\bf k}^*$ (and $\bar{\lambda}_{\bf k}^*$). 
Another important consequence is that the thermodynamical stability condition 
requires a constraint of the scalar off-equilibrium thermodynamical 
quantities  $\Lambda$, $\delta n$, 
and $\Pi$. 
This conditions practically determine both $\chi_+$ and $\chi_-$ 
(and thus also $\chi$), which are contributions neglected by the 
termination of polynomial function in the 0-th order irreducible tensor. 
By combing with the `equation of state' for off-equilibrium state 
eq.(\ref{eq:off-equilibrium_EoS}), 
the thermodynamical stability condition eq.(\ref{eq:stability}) 
also gives the differential equation eq.(\ref{eq:differential_fq_for_lambda}) for $\lambda_{\bf k}^*$ 
except for uncertainty in the integration constant $C_{\gamma}$.
As seen in the solutions to the differential equations,  
one can observe that $\chi_+$ and $\chi_-$, which can be fixed by the 
thermodynamical stability condition, play an important role in the 
energy dependence of $\lambda_{\bf k}^*$. 
Note that, the integration constant $C_{\gamma}$ defines a scale of 
$\lambda_{\bf k}^*$, 
or in other words, 
the constant $C_{\gamma}$ regulates the `strength' of 
the off-equilibrium state. This constant can be fixed by 
an index $\gamma\equiv \Pi/P_{\rm eq}$. 

Although $\alpha$, $\beta$, and $\gamma$ are to be obtained as 
solution of eq.(\ref{eq:formulation_summary}), 
we have investigated $\gamma$ dependence on the 
separation temperature $T$ and $\mu$ in a case that 
$u_{\mu}T^{\mu\nu} u_{\nu}$ and $N^{\mu} u_{\mu}$ are fixed 
and with varying $-\frac{1}{3} \Delta_{\mu\nu} T^{\mu\nu}$. 
In the case, the separation temperature $T$ and 
chemical potential $\mu$ obtained 
show (see Fig.\ref{fig:1} and \ref{fig:2}) that $T\approx T_0$ and 
$\mu\approx\mu_0$ hold for $\gamma < 10^{-3}$. 
In this region, the Landau matching condition 
approximately holds true. However, for $\gamma > 10^{-3}$, 
$T\ne T_0$ and $\mu\ne\mu_0$, i.e., the Landau matching condition is inappropriate and
its modification is required.

In summary, although the Landau matching condition 
determines the matching (fitting) temperature and 
chemical potential by using off-equilibrium energy density 
and baryon number density, these are not sufficient to 
specify the off-equilibrium state. 
In this article, we have introduced an index, 
namely, the strength of the off-equilibrium state 
$\gamma\equiv\Pi/P_{\rm eq}(T,\mu)$ in order to specify 
the state considered more precisely. 
The bulk pressure $\Pi$ may relate to the disturbance flow in a fluid cell 
and approaches to zero 
when the system relaxes to the local equilibrium state. 
Since the fluid dynamics is a deterministic theory, 
to set an initial condition of the system is crucially important 
and it should be done very carefully. 
In the literature to date, a constitutive equation 
for the bulk pressure is solved using a likely value under the Landau matching condition 
($\Lambda \equiv 0$ and $\delta n\equiv 0$). 
However, it must be inaccurate. 
The initial conditions (separation temperature $T$, 
chemical potential $\mu$ and bulk pressure $\Pi$) 
should be determined from the given off-equilibrium energy 
density $\varepsilon=u_{\mu}T^{\mu\nu}u_{\nu}$, 
net baryon number density 
$n=N^{\mu}u_{\mu}$, 
and the dissipative strength $\gamma$, 
as proposed in this article, i.e., eq.(\ref{eq:formulation_summary}).  
\section*{Acknowledgment}
The author would like to thank 
Prof. Grzegorz Wilk and Prof. Takeshi Kodama 
for helpful comments and discussions. 
The author acknowledges U. Heinz for crucial criticism concerning the 
original idea of this paper, first presented at the 9th Relativistic Aspects 
of Nuclear Physics (RANP 2013) conference at Rio de Janeiro, Brazil.
\appendix 
\section{Projection tensor}\label{app:A} 
The deviation from equilibrium $\phi(k,x)$ 
can be expanded by a series of irreducible tensors,  
\{$1$, $k^{\langle \mu \rangle}$, $k^{\langle\mu}k^{\nu\rangle}$, 
$k^{\langle\mu}k^{\nu}k^{\lambda\rangle}$, $\cdots$\},  forming a 
a complete and orthogonal set. 
One of components of the set of irreducible tensors is defined by
\begin{eqnarray}
   k^{\langle \mu_1} k^{\mu_2} \cdots k^{\mu_m\rangle} \equiv 
   \Delta^{\mu_1\mu_2\cdots \mu_m}_{\nu_1\nu_2\cdots \nu_m}~
   k^{\nu_1}k^{\nu_2}\cdots k^{\nu_m}, 
\end{eqnarray}
where a projection tensor, 
$\Delta^{\mu_1\mu_2\cdots \mu_m}_{\nu_1\nu_2\cdots \nu_m}$,
is given by (See also Appendix F in Ref.\cite{PhysRevD.85.114047}) 
\begin{subequations}
\begin{eqnarray}
\Delta^{\mu_1,\cdots\mu_m}_{\nu_1\cdots\nu_m} 
  = g_{\nu_1\sigma_1} \cdots g_{\nu_m\sigma_m}
    \Delta^{(\mu_1,\cdots\mu_m)(\sigma_1\cdots\sigma_m)}\quad 
\end{eqnarray}
and 
\begin{eqnarray}
   \Delta^{(\mu_1,\cdots\mu_m)(\nu_1\cdots\nu_m)} \equiv  
         \sum_{k=0}^{[m/2]} {\cal C}_{mk}~ 
       \Phi^{(\mu_1\cdots\mu_m)(\sigma_1\cdots\sigma_m)}_{mk}. 
\label{eq:def_irreducible_tensor}
\end{eqnarray}
\end{subequations}
The parentheses in the indexes of 
$\Phi^{(\mu_1\cdots\mu_m)(\sigma_1\cdots\sigma_m)}_{mk}$ 
denotes symmetrization 
under the exchange of indexes within $\mu_1\cdots\mu_m$ and 
 $\sigma_1\cdots\sigma_m$ of the tensor $\Phi_{mk}$. 
It is written by a sum of all possible permutations 
of $\mu$-type and $\nu$-type indexes as follows: 
\begin{eqnarray}  
  \Phi^{(\mu_1\cdots\mu_m)(\nu_1\cdots\nu_m)}_{mk}
  &\equiv&  
  \frac{1}{{\cal N}_{m,k}} \sum_{\wp_{\mu}, \wp_{\nu}}   
  \Delta^{\mu_1\mu_2} \cdots \Delta^{\mu_{2k-1}\mu_{2k}} \nonumber \\
  && \quad 
  \times 
  \Delta^{\nu_1 \nu_2} \cdots \Delta^{\nu_{2k-1}\nu_{2k}}  \times 
  \Delta^{\mu_{2k+1}\nu_{2k+1}}\cdots  \Delta^{\mu_{m}\nu_{m}}, 
  \label{eq:Phi_1}
\end{eqnarray} 
where 
$\sum_{\wp_{\mu}, \wp_{\nu}}$ 
represents the summation of all distinct permutations of the $\mu$-type and $\nu$-type
indexes. 
The coefficient ${\cal C}_{m,k}$ in eq.(\ref{eq:def_irreducible_tensor}) is 
introduced to satisfy the following conditions 
\begin{eqnarray}
  g_{\mu_i\mu_j} \Delta^{\mu_1\mu_2\cdots\mu_m}_{\nu_1\nu_2\cdots\nu_m} =0, \quad 
  g^{\nu_i\nu_j} \Delta^{\mu_1\mu_2\cdots\mu_m}_{\nu_1\nu_2\cdots\nu_m} =0, 
\end{eqnarray}
for $1\le i,j \le m$. 
The symbol $[m/2]$ denotes the largest integer not exceeding $m/2$, 
and the factor ${\cal N}_{m,k}$  is the total number of distinct permutations 
in the indexes $\mu$ and $\nu$ to be summed. 
In order to consider the symmetric tensor, 
let us introduce the following notations for the $\mu$-type 
rank-2 and rank-4 tensors   
\begin{subequations}
\begin{eqnarray}
  \{ \Delta ^{(\mu_i\mu_j)}_{m} \}    
  &\equiv&
  \frac{1}{2} \sum_{\wp_{\mu}; i,j\le m}\Delta^{\mu_i\mu_j} = 
  \frac{1}{2} \sum_i^m \sum_{j\ne i}^m  \Delta^{\mu_i \mu_j}, \qquad 
  \label{eq:open_square1}\\
  \{\Delta ^{(\mu_i\mu_j)}_{m}\}^{2} 
  &\equiv&
  \frac{1}{2^2}\frac{1}{2!}  \sum_{\wp_{\mu};i,j,k,l \le m} \Delta^{\mu_i\mu_j}
\Delta^{\mu_k\mu_l} 
  \nonumber \\ 
&=&\frac{1}{2^2}\frac{1}{2!} 
  \sum_i^m \sum_{j\ne i}^m \sum_{k \ne i\ne j}^m \sum_{l\ne i\ne j\ne k}^m
  \Delta^{\mu_i \mu_j} \Delta^{\mu_k \mu_l},  
  \label{eq:open_square2} 
\end{eqnarray}
\end{subequations} 
respectively, where 
the factors $\frac{1}{2}$ and $\frac{1}{2^2}\frac{1}{2!}$ 
in eq.(\ref{eq:open_square1}) and eq.(\ref{eq:open_square2})  
are introduced so as to exclude duplication caused by the possible 
exchange of suffixes. 
For example, $\{\Delta^{(\mu_i\mu_j)}_{m}\}^{k}$ with $m=4$ and $k=2$ 
means the following rank-$2k$ tensor, 
\begin{eqnarray}
 && \{\Delta^{(\mu_i\mu_j)}_{4}\}^{2} =  
 \Delta^{\mu_1\mu_2} \Delta^{\mu_3\mu_4} + 
 \Delta^{\mu_1\mu_3} \Delta^{\mu_2\mu_4} +
 \Delta^{\mu_1\mu_4} \Delta^{\mu_2\mu_3}. \nonumber 
\end{eqnarray}
Note that, possible permutations of the suffixes include 
\begin{eqnarray*}
 \Delta^{\mu_1\mu_2} \Delta^{\mu_4\mu_3},~
 \Delta^{\mu_2\mu_1} \Delta^{\mu_3\mu_4},~
 \Delta^{\mu_2\mu_1} \Delta^{\mu_4\mu_3}, 
\end{eqnarray*}
which gives exactly the same contribution as the above first term 
$\Delta^{\mu_1\mu_2} \Delta^{\mu_3\mu_4}$. In addition to this, 
the permutation also includes terms such as 
$\Delta^{\mu_3\mu_4}\Delta^{\mu_1\mu_2}$, which is also gives the 
same contribution. In order to exclude these terms, the factor 
$(\frac{1}{2})^k \frac{1}{k!}$ is needed. 
Thus, for the general $2k$-rank tensor, 
$\{ \Delta ^{(\mu_i\mu_j)}_{m} \}^k$ ($k\le m/2$), 
it is defined by 
\begin{subequations}\label{eq:non-mixed-type}  
\begin{eqnarray}
\{ \Delta ^{(\mu_i\mu_j)}_{m} \}^k 
&\equiv&
 \frac{1}{2^k}\frac{1}{k!}  \sum_{\wp_{\mu}}
 \Delta^{\mu_{i_1} \mu_{i_2}}  
  \cdots \Delta^{\mu_{2k-1} \mu_{2k}}, \quad 
\end{eqnarray} 
where $m$ denotes a permissible maximum number of the suffix.  
Similarly for $\nu$-type tensors, we define 
\begin{eqnarray}
\{ \Delta ^{(\nu_i\nu_j)}_{m} \}^k 
&\equiv&
 \frac{1}{2^k}\frac{1}{k!}  \sum_{\wp_{\nu}}
 \Delta^{\nu_{i_1} \nu_{i_2}}  
  \cdots \Delta^{\nu_{2k-1} \nu_{2k}}.\quad 
\end{eqnarray}
\end{subequations}
Since the tensor  $\Phi^{(\mu_1\cdots\mu_m)(\sigma_1\cdots\sigma_m)}_{mk}$ 
is symmetric under permutations of both $\mu$ and $\nu-$type indexes, 
the last factor in the second line of eq.(\ref{eq:Phi_1}) 
such as $\Delta^{\mu_{2k+1}\nu_{2k+1}}\cdots  \Delta^{\mu_{m}\nu_{m}}$ is 
different from other factors of eq.(\ref{eq:Phi_1}) denoting 
$\{\Delta ^{(\mu_i\mu_j)}_{m} \}^k$ and $\{\Delta ^{(\nu_i\nu_j)}_{m} \}^k$.   
Let us introduce a similar representation 
for a different symmetry as shown in the following equation;  
\begin{eqnarray}
\{ \Delta ^{\mu_i\nu_j}_{m}\}^k \equiv 
\frac{1}{k!} \sum_{\wp_{\mu}}\sum_{\wp_{\nu}}  
\Delta^{\mu_{i_1} \nu_{j_1}} \cdots \Delta^{\mu_{i_k} \nu_{j_k}}. 
 \label{eq:mixed-type}   
\end{eqnarray} 
Then, using these definitions given in eqs.(\ref{eq:non-mixed-type}) and (\ref{eq:mixed-type}), 
we express the tensor eq.(\ref{eq:Phi_1}) as
\begin{eqnarray}
&&
\Phi^{(\mu_1\cdots\mu_m)(\nu_1\cdots\nu_m)}_{mk} 
 = \frac{1}{{\cal N}_{m,k}} 
    \{ \Delta^{(\mu_i\mu_j)}_{m} \}^{k}
    \{\Delta^{(\nu_i\nu_j)}_{m} \}^{k}  
    \{ \Delta^{(\mu_l\nu_l)}_{m-2k} \}^{m-2k}. 
\end{eqnarray} 
We hereafter denote $n^{(\mu_i\mu_j)}_{m,k}$, 
$n^{(\nu_i\nu_j)}_{m,k}$ and $n^{(\mu_l\nu_l)}_{m,k}$ 
as the total number of distinct permutations 
contained in $\{ \Delta^{(\mu_i\mu_j)}_{m} \}^{k}$, 
$\{ \Delta^{(\nu_i\nu_j)}_{m} \}^{k}$ and 
$\{ \Delta^{(\mu_l\nu_l)}_{m} \}^{k}$, respectively. 
Thus, we have 
\begin{subequations}
\begin{eqnarray}
&&
 n^{(\mu_i\mu_j)}_{m,k}= n^{(\nu_i\nu_j)}_{m,k} = 
  \frac{{}_{m} P_{2k}}{k! 2^k},  \\ 
&&   n^{(\mu_l\nu_l)}_{m,k} = 
  \frac{ [{}_{m} P_{k}]^2}{k!}  = \frac{[m!]^2}{[(m-k)!]^2 k!}. 
\end{eqnarray} 
\end{subequations} 
${\cal N}_{m,k}$ is the total number of possible permutations belonging 
to the projection tensor 
$\Phi^{(\mu_1\cdots\mu_m)(\nu_1\cdots\nu_m)}_{mk}$, which is given by 
\begin{eqnarray}
 {\cal N}_{m,k} &=&
   n^{(\mu_i\mu_j)}_{m,k} ~n^{(\nu_i\nu_j)}_{m,k}  ~n^{(\mu_l\nu_l)}_{m-2k,m-2k}
 \nonumber \\ 
&=& \left[ \frac{m!}{k!2^k} \right]^2\frac{1}{(m-2k)!}  ~. 
\end{eqnarray}  
Considering linear combinations for the projection tensor 
$\Phi^{(\mu_1\cdots\mu_m)(\nu_1\cdots\nu_m)}_{mk}$ 
with different $k$ values 
we obtain 
\begin{eqnarray}
    \Delta^{(\mu_1\cdots\mu_m) (\nu_1\cdots\nu_m)} \equiv 
    \sum_{k=0}^{[m/2]} \frac{{\cal C}_{m,k}}{{\cal N}_{m,k}}
    \{ \Delta^{(\mu_i\mu_j)}_{m} \}^{k}
    \{\Delta^{(\nu_i\nu_j)}_{m} \}^{k}  
    \{ \Delta^{(\mu_l\nu_l)}_{m-2k} \}^{m-2k}. \qquad 
\end{eqnarray}
Note that, the value of $C_{m,0}$ can be set arbitrarily 
so that it may be set to the value of  1 
without loss of generality. Thus, 
${\cal C}_{m,k}$ can then be determined as the following: 
\begin{eqnarray}
{\cal C}_{m,k} = (-1)^k~\frac{(m!)^2}{(2m)!} ~\frac{(2m-2k)!}{k!(m-k)!(m-2k)!}~.
\label{eq:Cmk}  
\end{eqnarray} 
The derivation of eq.(\ref{eq:Cmk}) is given in Appendix \ref{app:B}.

\section{Derivation of the ${\cal C}_{m,k}$}\label{app:B} 
When the indices $\mu_1$ and $\mu_2$ are expressed explicitly, 
the projection tensor is written as follows: 
\begin{eqnarray*}
\Delta^{(\mu_1\mu_2\cdots\mu_m) (\nu_1\cdots\nu_m)} 
&=&\sum_{k=0}^{[m/2]} \frac{{\cal C}_{m,k}}{{\cal N}_{m,k}} \Big[  
  \Delta^{\mu_1\mu_2} 
  \{\Delta^{(\mu_i\mu_j)}_{m-2} \}^{k-1} 
  \{\Delta^{(\nu_i\nu_j)}_{m} \}^{k}  
  \{ \Delta^{(\mu_l\nu_l)}_{m-2k} \}^{m-2k} 
\\ 
&&\qquad 
+ 
  \{ \Delta^{(\mu_1\mu_{i'})}_{m-2}   \} 
  \{ \Delta^{(\mu_2\mu_{j'})}_{m-3}   \}
  \{ \Delta^{(\mu_{i}\mu_{j})}_{m-4}\}^{k-2} 
  \{ \Delta^{(\nu_i\nu_j)}_{m} \}^{k}  
  \{ \Delta^{(\mu_l\nu_l)}_{m-2k} \}^{m-2k} 
\\ 
&&\qquad 
+  
  \{ \Delta^{(\mu_1 \mu_{i'})}_{m-2} \} 
  \{ \Delta^{(\mu_i\mu_j)}_{m-3} \}^{k-1} 
  \{ \Delta^{(\nu_i\nu_j)}_{m} \}^{k}   
  \{ \Delta^{(\mu_2 \nu_{i'})}_{m-2k} \} 
  \{ \Delta^{(\mu_{l}\nu_{l})}_{m-2k-1} \}^{m-2k-1} \\   
&&\qquad +  
  \{ \Delta^{(\mu_2 \mu_{i'})}_{m-2} \} 
  \{ \Delta^{(\mu_i\mu_j)}_{m-3} \}^{k-1} 
  \{ \Delta^{(\nu_i \nu_j)}_{m} \}^{k}   
  \{ \Delta^{(\mu_1\nu_{i'})}_{m-2k} \} 
  \{ \Delta^{(\mu_{l}\nu_{l})}_{m-2k-1} \}^{m-2k-1} \\   
&&\qquad +  
  \{ \Delta^{(\mu_i\mu_j)}_{m-2} \}^{k}
  \{ \Delta^{(\nu_i\nu_j)}_{m} \}^{k} 
  \{ \Delta^{(\mu_1\nu_{i'})}_{m-2k} \} 
  \{ \Delta^{(\mu_2 \nu_{j'})}_{m-2k-1} \} 
  \{ \Delta^{(\mu_l\nu_l)}_{m-2k-2} \}^{m-2k-2} 
 ~\Big]. 
\end{eqnarray*}
Here, we take contraction of two indices $\mu_1$ and $\mu_2$ and obtain 
\begin{eqnarray}
g_{\mu_1\mu_2} 
    \Delta^{(\mu_1\cdots\mu_m) (\nu_1\cdots\nu_m)}
&=& \sum_{k=0}^{[m/2]} \frac{{\cal C}_{m,k}}{{\cal N}_{m,k}} 
    \Big[ 
      ~x  \{\Delta^{(\mu_i\mu_j)}_{m-2} \}^{k-1}
       \{\Delta^{(\nu_i\nu_j)}_{m} \}^{k}  
       \{ \Delta^{(\mu_l\nu_l)}_{m-2k} \}^{m-2k}  \nonumber \\ 
&&
       +c^{(1)}_{m,k} 
       \{\Delta^{(\mu_i\mu_j)}_{m-2}\}^{k-1}
       \{\Delta^{(\nu_i\nu_j)}_{m} \}^{k}  
       \{ \Delta^{(\mu_l\nu_l)}_{m-2k} \}^{m-2k} \nonumber \\
&& + 2c^{(2)}_{m,k}
      \{ \Delta^{(\mu_i\mu_j)}_{m-2} \}^{k-1}  
      \{\Delta^{(\nu_i\nu_j)}_{m} \}^{k}  
      \{ \Delta^{(\mu_l\nu_l)}_{m-2k} \}^{m-2k} \nonumber \\
&&
      +c^{(3)}_{m,k}
      \{ \Delta^{(\mu_i\mu_j)}_{m-2} \}^{k}  
      \{\Delta^{(\nu_i\nu_j)}_{m} \}^{k+1}  
      \{ \Delta^{(\mu_l\nu_l)}_{m-2k-2} \}^{m-2(k+1)}  
      \Big] =0, \quad \label{eq:trace=0}
\end{eqnarray}
where $x\equiv g_{\mu_1\mu_2} \Delta^{\mu_1\mu_2}$ and 
$c^{(1)}_{mk},c^{(2)}_{mk},c^{(3)}_{mk}$ are given by 
\begin{subequations}
\begin{eqnarray}
c^{(1)}_{m,k} &=& \frac{
 n^{(\mu_1\mu_{i'})}_{m-2,1}~ n^{(\mu_2\mu_{j'})}_{m-3,1}~ 
 n^{(\mu_i\mu_j)}_{m-4,k-2} 
  }{
  n^{(\mu_i\mu_j)}_{m-2,k-1}
  }= 2k-2, \qquad \\ 
c^{(2)}_{m,k}&=& \frac{
  n^{(\mu_1\mu_{i'})}_{m-2,1}~
  n^{(\mu_i\mu_j)}_{m-3,k-1}~
  n^{(\mu_2\nu_{i'})}_{m-2k,1}
}
{
 n^{(\mu_i\mu_j)}_{m-2,k-1}
} 
~\frac{
  n^{(\mu_l\nu_l)}_{m-2k-1,m-2k-1}
}{
  n^{(\mu_l\nu_l)}_{m-2k,m-2k}
}
=m-2k,\qquad \\
c^{(3)}_{m,k}&=&
\frac{
   n^{(\nu_i\nu_j)}_{m,k}
   ~n^{(\mu_1\nu_{i'})}_{m-2k,1}
   ~n^{(\mu_2\nu_{i'})}_{m-2k-1,1} 
}{ 
   n^{(\nu_i\nu_j)}_{m,k+1}   
} 
~\frac{ 
    n^{(\nu_i\nu_j)}_{m-2k-2,m-2k-2}
}{ 
   n^{(\mu_i\nu_j)}_{m-2k-2,m-2k-2}
}
= 2k+2. \qquad 
\end{eqnarray}
\end{subequations}
Since the contraction $g_{\mu_1\mu_2} 
\Delta^{(\mu_1\cdots\mu_m) (\nu_1\cdots\nu_m)} = 0$, 
the coefficients $c^{(1)}_{m,k}$, $c^{(2)}_{m,k}$ and $c^{(3)}_{m,k}$ 
need to satisfy the following equation: 
\begin{eqnarray}
    \frac{{\cal C}_{m,k-1}}{{\cal N}_{m,k-1}} c^{(3)}_{m,k-1} 
  +\frac{{\cal C}_{m,k}}{{\cal N}_{m,k}} 
   (x+c^{(1)}_{m,k}+2c^{(2)}_{m,k}) =0. \quad  
   \label{eq:pre_recursion_formula}
\end{eqnarray} 
From the above equation, we obtain 
the following recursion formula for $ {\cal C}_{m,k}$; 
\begin{eqnarray}
  {\cal C}_{m,k}= -\frac{(m-2k+2)(m-2k+1)}{2k(2m-2k+1)} ~{\cal C}_{m,k-1} . \quad 
\end{eqnarray}
Since the coefficient $C_{m,0}$ can be set to 1, 
${\cal C}_{m,k}$ for general $k$ is determined according to 
\begin{eqnarray}
{\cal C}_{m,k} = (-1)^k~\frac{(m!)^2}{(2m)!} ~\frac{(2m-2k)!}{k!(m-k)!(m-2k)!}~, 
\end{eqnarray} 
which is eq.(\ref{eq:Cmk}). 

\section{Total contraction of the projection tensor}\label{app:C} 
An identity 
\begin{eqnarray*}
  g_{\mu_1\nu_1}\cdots g_{\mu_m\nu_m} 
  \Delta^{(\mu_1\cdots\mu_m)(\nu_1\cdots\nu_m)} = 2m+1,\qquad 
\end{eqnarray*}
will be used in derivation of the orthogonality condition for the irreducible 
projection tensors in the later Appendix \ref{app:D}. Let us proof the identity in this section.  
  
When the indices $\mu_1$ and $\nu_1$ are expressed explicitly, 
the projection tensor is written as follows:  
\begin{eqnarray*} 
\Delta^{(\mu_1\cdots\mu_m)(\nu_1\cdots\nu_m)} 
&=& \sum_{k=0}^{[m/2]} \frac{{\cal C}_{m,k}}{{\cal N}_{m,k}} \Big[  
  \{\Delta^{(\mu_1\mu_{i'})}_{m-1}\} 
  \{\Delta^{(\mu_i\mu_j)}_{m-2} \}^{k-1} 
  \{\Delta^{(\nu_1\nu_{i'})}_{m-1} \}
  \{\Delta^{(\nu_i\nu_j)}_{m-2} \}^{k-1}
  \{ \Delta^{(\mu_{l}\nu_{l})}_{m-2k} \}^{m-2k} \\ 
&&\qquad + 
  \{ \Delta^{(\mu_i\mu_j)}_{m-1}\}^k    
  \{ \Delta^{(\nu_1\nu_{i'})}_{m-1} \}
  \{ \Delta^{(\nu_i\nu_j)}_{m-2} \}^{k-1}
  \{ \Delta^{(\mu_1\nu_{l'})}_{m-2k} \}   
  \{ \Delta^{(\mu_{l}\nu_{l})}_{m-2k-1} \}^{m-2k-1} \\ 
&&\qquad + 
  \{ \Delta^{(\mu_1\mu_{i'})}_{m-1}\}
  \{ \Delta^{(\mu_i\mu_j)}_{m-2}\}^{k-1} 
  \{ \Delta^{(\nu_i\nu_j)}_{m-1} \}^{k} 
  \{ \Delta^{(\mu_{l'}\nu_1)}_{m-2k} \}   
  \{ \Delta^{(\mu_l\nu_l)}_{\in m-2k-1} \}^{m-2k-1} \\ 
&&\qquad + 
  \{ \Delta^{(\mu_i\mu_j)}_{m-1}\}^{k}  
  \{ \Delta^{(\nu_i\nu_j)}_{m-1} \}^{k} 
  ~\Delta^{\mu_1\nu_1}~
  \{ \Delta^{(\mu_l\nu_l)}_{m-2k-1} \}^{m-2k-1} \\ 
&&\qquad +  
  \{ \Delta^{(\mu_i\mu_j)}_{m-1}\}^{k}  
  \{ \Delta^{(\nu_i\nu_j)}_{m-1} \}^{k} 
  \{ \Delta^{(\mu_{l'}\nu_1)}_{m-2k-1}\}   
  \{ \Delta^{(\mu_1\nu_{l'})}_{m-2k-1} \}
  \{ \Delta^{(\mu_l\nu_l)}_{m-2k-2} \}^{m-2k-2} \Big]. 
\end{eqnarray*}
Contracting the indices $\mu_1$ and $\nu_2$, we get 
\begin{eqnarray}
 && g_{\mu_1\nu_1} 
  \Delta^{(\mu_1\cdots\mu_m)(\nu_1\cdots\nu_m)} = 
  \sum_{k=0}^{[m/2]} \frac{{\cal C}_{m,k}}{{\cal N}_{m,k}} \Big[ 
  d^{(1)}_{m,k}~ 
  \{ \Delta^{(\mu_i\mu_j)}_{m-1} \}^{k-1} 
  \{ \Delta^{(\nu_i\nu_j)}_{m-1} \}^{k-1} 
  \{ \Delta^{(\mu_l\nu_l)}_{m-2k+1} \}^{m-2k+1} \nonumber \\ 
&&\quad +  
    [ d^{(2)}_{m,k} + d^{(3)}_{m,k} + x d^{(4)}_{m,k} + d^{(5)}_{m,k} ]~ 
 \{ \Delta^{(\mu_i\mu_j)}_{m-1} \}^{k} 
 \{ \Delta^{(\nu_i\nu_j)}_{m-1} \}^{k} 
 \{ \Delta^{(\mu_l\nu_l)}_{m-2k-1} \}^{m-2k-1} \Big],  
\end{eqnarray}
where 
\begin{subequations}
\begin{eqnarray}
d^{(1)}_{m,k} &\equiv&
\frac{
   n^{(\mu_1\mu_{i'})}_{m-1,1}
  ~n^{(\mu_i\mu_j)}_{m-2,k-1}
  ~n^{(\nu_1\nu_{i'})}_{m-1,1} 
  ~n^{(\nu_i\nu_j)}_{m-2,k-1}
  ~n^{(\mu_l\nu_l)}_{m-2k,m-2k}
}{
  n^{(\mu_i\mu_j)}_{m-1,k-1} 
 ~n^{(\nu_i\nu_j)}_{m-1,k-1}
 ~n^{(\mu_l\nu_l)}_{m-2k+1,m-2k+1}
}= m-2k+1, \\ 
d^{(2)}_{m,k} &\equiv&
\frac{
  n^{(\mu_i\mu_j)}_{m-1,k} 
 ~n^{(\nu_1\nu_{i'})}_{m-1,1}
 ~n^{(\nu_i\nu_j)}_{m-2,k-1} 
 ~n^{(\mu_1\nu_{l'})}_{m-2k,1} 
 ~n^{(\mu_l\nu_l)}_{m-2k-1,m-2k-1}
}{
  n^{(\mu_i\mu_j)}_{m-1,k}
 ~n^{(\nu_i\nu_j)}_{m-1,k}
 ~n^{(\mu_l\nu_l)}_{m-2k-1,m-2k-1}
}= 2k,\\ 
d^{(3)}_{m,k} &\equiv&
\frac{
  n^{(\mu_1\mu_{i'})}_{m-1,1} 
 ~n^{(\mu_i\mu_j)}_{m-2,k-1}
 ~n^{(\nu_i\nu_j)}_{m-1,k} 
 ~n^{(\mu_{i'}\nu_1)}_{m-2k,1} 
 ~n^{\mu_l\nu_l}_{m-2k-1,m-2k-1} 
}{
  n^{(\mu_i\mu_j)}_{m-1,k} 
 ~n^{(\nu_i\nu_j)}_{m-1,k} 
 ~n^{(\mu_l\nu_l)}_{m-2k-1,m-2k-1}  
}= 2k,\\ 
d^{(4)}_{m,k} &\equiv&
\frac{
  n^{(\mu_i\mu_j)}_{m-1,k}
 ~n^{(\nu_i\nu_j)}_{m-1,k}
 ~n^{(\mu_l\nu_l)}_{m-2k-1,m-2k-1} 
}{
   n^{(\mu_i\mu_j)}_{m-1,k} ~n^{(\nu_i\nu_j)}_{m-1,k}~n^{(\mu_l\nu_l)}_{m-2k-1,m-2k-1} 
}=1,\\ 
d^{(5)}_{m,k} &\equiv&
\frac{
  n^{(\mu_i\mu_j)}_{m-1,k} 
 ~n^{(\nu_i\nu_j)}_{m-1,k} 
 ~n^{(\mu_{i'}\nu_1)}_{m-2k-1,1}
 ~n^{(\mu_1\nu_{i'})}_{m-2k-1,1} 
 ~n^{(\mu_l\nu_l)}_{m-2k-2,m-2k-2}
}{
  n^{(\mu_i\mu_j)}_{m-1,k}
 ~n^{(\nu_i\nu_j)}_{m-1,k} 
 ~n^{(\mu_l\nu_l)}_{m-2k-1,m-2k-1}  
}=m-2k+1. \quad 
\end{eqnarray}
\end{subequations} 
Hence, we obtain a formula 
\begin{eqnarray}
 g_{\mu_1\nu_1} \Delta^{(\mu_1\cdots\mu_m)(\nu_1\cdots\nu_m)} 
=\sum_{k=0}^{[m/2]}
 \gamma_{m,k}
 \Big[~ 
  (m-2k+1)~{\cal M}_{m-1,k-1} 
  +( m+2k+2 )~{\cal M}_{m-1,k} 
\Big], ~
\end{eqnarray}
where 
\begin{eqnarray}
  \gamma_{m,k} &\equiv& \frac{{\cal C}_{m,k}}{{\cal N}_{m,k}}, \\ 
  {\cal M}_{m,k} &\equiv& 
 \{ \Delta^{(\mu_i\mu_j)}_{m} \}^{k} 
 \{ \Delta^{(\nu_i\nu_j)}_{m} \}^{k} 
 \{ \Delta^{(\mu_l\nu_l)}_{m-2k} \}^{m-2k}.
\end{eqnarray}
This means 
\begin{eqnarray}
 g_{\mu_1\nu_1} \Delta^{(\mu_1\cdots\mu_m)(\nu_1\cdots\nu_m)}  
   &=& \frac{2m+1}{2m-1} \sum_{k=0}^{[m/2]-1} \gamma_{m-1,k} {\cal M}_{m-1,k}
\nonumber \\ 
  &=& \frac{2m+1}{2m-1}
       \Delta^{(\mu_2\cdots\mu_{m})(\nu_2\cdots\nu_{m})}. 
\end{eqnarray}
Using the above formula iteratively, we obtain 
\begin{eqnarray}
  g_{\mu_1\nu_1}\cdots g_{\mu_m\nu_m} 
  \Delta^{(\mu_1\cdots\mu_m)(\nu_1\cdots\nu_m)} = 2m+1. \qquad 
  \label{eq:2m+1} 
\end{eqnarray} 
\section{Derivation of the orthogonality condition}\label{app:D}
In order to discuss the orthogonality of the irreducible projection tensor, 
 consider the following integral 
for the case that $F_{\bf k}$ is an arbitrary function of 
$E_k\equiv u_{\mu}k^{\mu}$ only; 
\begin{eqnarray}
 {\cal J}_m &=& \int\! \frac{d^3{\bf k} ~F_{\bf k}}{(2\pi)^3k^0}~ 
 k^{\langle \mu_1}k^{\mu_2}\cdots k^{\mu_m\rangle} 
 k_{\langle \nu_1}k_{\nu_2}\cdots k_{\nu_m\rangle} \nonumber  \\ 
&=& \int\! \frac{d^3k~F_{\bf k}}{(2\pi)^3k^0}~   [k^{\lambda}k_{\lambda}]^m 
  ~g^{\alpha_1}_{\beta_1}  
  \cdots g^{\alpha_m}_{\beta_m}
  ~\Delta^{\mu_1\cdots\mu_m}_{\alpha_1\cdots\alpha_m}
  ~\Delta^{\beta_1\cdots\beta_m}_{\nu_1\cdots\nu_m} \nonumber \\ 
&=& \Delta^{\mu_1\cdots\mu_m}_{\nu_1\cdots\nu_m} 
  \int\! \frac{d^3k~F_{\bf k}}{(2\pi)^3k^0}~   [k^{\lambda}k_{\lambda}]^m   
    \frac{ g^{\alpha_1}_{\beta_1}  
   \cdots g^{\alpha_m}_{\beta_m}  }{2m+1} 
  \Delta^{\beta_1\cdots\beta_m}_{\alpha_1\cdots\alpha_m}
  \nonumber \\
&=& ~\frac{\Delta^{\mu_1\cdots\mu_m}_{\nu_1\cdots\nu_m}}{2m+1}  
    \int\! \frac{d^3k~F_{\bf k}}{(2\pi)^3k^0}~
   k^{\langle \alpha_1}\cdots k^{\alpha_m\rangle} 
   k_{\langle \alpha_1}\cdots k_{\alpha_m\rangle},
   \label{eq:other_than_2m+1-0}  
\end{eqnarray} 
where we require 
\begin{eqnarray}
  \Delta^{(\mu_1\cdots\mu_m)(\nu_1\cdots\nu_m)}
    =~ g_{\alpha_1\beta_1} 
  \cdots g_{\alpha_m\beta_m}
  \Delta^{(\mu_1\cdots\mu_m)(\alpha_1\cdots\alpha_m)}
  \Delta^{(\beta_1\cdots\beta_m)(\nu_1\cdots\nu_m)}, 
\end{eqnarray}
and eq.(\ref{eq:2m+1}). 
The integral part in eq.(\ref{eq:other_than_2m+1-0}) 
can be evaluated as 
\begin{eqnarray}
&& \int\! \frac{d^3k~F_k}{(2\pi)^3k^0}
   k^{\langle \alpha_1}\cdots k^{\alpha_m\rangle } 
   k_{\langle \alpha_1}\cdots k_{\alpha_m\rangle } 
   =\sum_{j=0}^{[m/2]} \gamma_{m,j}\times {\cal N}_{m,j}
      \int\! \frac{d^3k~F_k}{(2\pi)^3k^0}   
    [\Delta_{\alpha\beta}k^{\alpha}k^{\beta}]^m \nonumber \\ 
&&\quad = \frac{(m!)^2 2^m}{(2m)!} P_m(1) \int\! \frac{d^3k~F_k}{(2\pi)^3k^0} 
         [\Delta_{\alpha\beta}k^{\alpha}k^{\beta}]^m 
  =  \frac{m!}{(2m-1)!!} ~\int\! \frac{d^3k~F_k}{(2\pi)^3k^0} ~
         [\Delta_{\alpha\beta}k^{\alpha}k^{\beta}]^m, \label{eq:other_than_2m+1} \qquad 
\end{eqnarray}
where $P_m(x)$ is a Legendre function of the order of $m$. 
Hence, we can finally obtain 
\begin{eqnarray}
&& \int\! \frac{d^3k~F_k}{(2\pi)^3k^0}~ 
 k^{\langle \mu_1}k^{\mu_2}\cdots k^{\mu_m\rangle} 
 k_{\langle \nu_1}k_{\nu_2}\cdots k_{\nu_m\rangle}  
 ~= \frac{m!}{(2m+1)!!}~\int\! \frac{d^3k~F_k}{(2\pi)^3k^0}
         [\Delta_{\alpha\beta}k^{\alpha}k^{\beta}]^m, \qquad 
\end{eqnarray}
which is eq.(\ref{eq:orthogonality_condition_final_b}).  
\section{Evaluation of the $I^{(l)}_r$}\label{app:E}
We require to calculate the following integrals denoting $ I^{(l)}_r$ 
\begin{eqnarray}
  I^{(l)}_r &\equiv& \int\!\frac{d^3{\bf k}~\omega^{(l)}_{\bf k}}
 {(2\pi)^3k_0} (E_{\bf k} )^r \nonumber \\ &=&
 \frac{(-1)^l}{2\pi^2} \frac{{\cal W}^{(l)}}{(2l+1)!!}
  \int\! \frac{k^2 dk}{E_{\bf k}}  \Big[ \sqrt{E_{\bf k}^2-m^2} \Big]^{2l}~ [E_{\bf k}]^r 
  ~e^{-(E_k-\mu_{\rm b})/T} .
\end{eqnarray} 
This can be expressed in the following form, 
\begin{eqnarray}
I^{(l)}_r &=& \frac{(-1)^l (-1)^r }{2\pi^2} 
   \frac{{\cal W}^{(l)}}{(2l+1)!!}~e^{\mu_{\rm b}/T}~ (aT)^{2l+2+r} 
   (\frac{d^r}{da^r})  
  \int_{0}^{\infty}\! dy~[\sinh y]^{2l+2}  ~e^{-a\cosh y} \nonumber \\ 
&=& \frac{(-1)^{l+r}}{2\pi^2} {\cal W}^{(l)}~e^{\mu_{\rm b}/T}~ (aT)^{2l+2+r}~ 
  \frac{d^r}{da^r}[ \frac{1}{a^{l+1} } K_{l+1}(a)], 
\end{eqnarray}
where $a=m/T$. 
Using the identity about the following differential operators we have 
\begin{eqnarray}
   \Big[ \frac{d}{dz} \Big]^r = \sum_{k=0}^{[r/2]} 
   \frac{r!}{2^k k!(r-2k)!} ~z^{(r-2k)} \Big[ \frac{1}{z} \frac{d}{dz} \Big]^{r-k}.
\end{eqnarray}
We then obtain
\begin{eqnarray}
  I^{(l)}_r 
&=& \frac{1}{2\pi^2} {\cal W}^{(l)} ~e^{\mu_{\rm b}/T}~T^{2(l+1)+r} 
   \times\sum_{k=0}^{[r/2]} 
   \frac{ (-1)^{l-k} ~r!}{2^k k!(r-2k)!} 
   ~a^{(l+1)+r-k} K_{(l+1)+r-k}(a). \qquad
\end{eqnarray}  
By requiring $I^{(l)}_0\equiv 1$, we can determine the normalization factor, 
\begin{eqnarray}
  1/{\cal W}^{(l)} = 
  \frac{(-1)^{l} }{2\pi^2} e^{\mu_{\rm b}/T}~T^{2(l+1)}~
    (\frac{m}{T})^{(l+1)} K_{(l+1)}(\frac{m}{T}).\qquad 
\end{eqnarray} 
Moreover, substituting the result obtained previously in the 
normalization factor ${\cal W}^{(l)}$, we finally obtain 
\begin{eqnarray}
 I^{(l)}_r &=& m^r r!  \sum_{k=0}^{[r/2]} 
   \frac{(-1)^{k}}{2^k k!(r-2k)!}~(\frac{m}{T})^{-k} 
    \frac{K_{l+1+r-k}(m/T)}{K_{l+1}(m/T)}.
\end{eqnarray}
\section{Derivation of eq.(\ref{eq:from_micro_to_macro1})}\label{app:F}
The coefficient tensors $\lambda_{\bf k}^{\langle \mu_1\cdots\mu_l\rangle} $ 
and $\bar{\lambda}_{\bf k}^{\langle \mu_1\cdots\mu_l\rangle}$ must be 
rewritten as linear combinations of $\rho^{\mu_1\cdots\mu_l}_m$ or 
$\bar{\rho}^{\mu_1\cdots\mu_l}_m$ using the expressions given in eq.(\ref{eq:cn2}). 
Thus, we resubstitute eq.(\ref{eq:cn2a}) into the definition 
of $\lambda_{\bf k}$ given by eq.(\ref{eq:lambda_k}) to yield the following 
expression
\begin{eqnarray}
\lambda_{\bf k}^{\langle \mu_1\cdots \mu_l \rangle} 
&=& \frac{{\cal W}^{(l)}}{l!}\left(
  \begin{array}{llll} 
  \rho_0^{\mu_1\cdots\mu_l},&
  \rho_1^{\mu_1\cdots\mu_l},&
  \cdots, &
  \rho_{N_l}^{\mu_1\cdots\mu_l}
  \end{array} 
   \right) \nonumber \\ 
&& \times  
  \left( \begin{array}{llll} 
  a^{(l)}_{00} & a^{(l)}_{10}  & \ldots & a^{(l)}_{N_l, 0}  \\ 
  0 & a^{(l)}_{11}  & \ldots & a^{(l)}_{N_l,1} \\ 
 \ldots        &\ldots & \ldots  &\ldots \\  
 0  &  0 &\ldots &  a^{(l)}_{N_l,N_l} \\ 
 \end{array} \right) 
  \left( \begin{array}{lllll} 
  a^{(l)}_{00} &0& \ldots & 0  \\ 
  a^{(l)}_{10} & a^{(l)}_{11}  & \ldots   & 0 \\ 
 \ldots     &\ldots     &\ldots   & 0 \\  
 a^{(l)}_{N_l,0}    & a^{(l)}_{N_l,1} & \ldots &  a^{(l)}_{N_l,N_l} \\ 
 \end{array} \right)
\left( \begin{array}{c} 
 E_{\bf k}^0 \\ 
 E_{\bf k}^1\\ \cdots \\  
 ~E_{\bf k}^{N_l}\\ 
 \end{array} \right),  \qquad
\end{eqnarray} 
where we use eq.(\ref{eq:Pn}) in the following matrix form
\begin{eqnarray}
\left( \begin{array}{c} 
 P_{{\bf k}0}^{(l)} \\ 
 P_{{\bf k}1}^{(l)}\\ \cdots \\  
 P_{{\bf k}n}^{(l)}\\ 
 \end{array} \right)=
  \left( \begin{array}{lllll} 
  a^{(l)}_{00} &0& \ldots & 0  \\ 
  a^{(l)}_{10} & a^{(l)}_{11}  & \ldots   & 0 \\ 
 \ldots     &\ldots     &\ldots   & 0 \\  
 a^{(l)}_{N_l,0}    & a^{(l)}_{N_l,1} & \ldots &  a^{(l)}_{N_l,N_l} \\ 
 \end{array} \right)
\left( \begin{array}{c} 
 E_{\bf k}^0 \\ 
 E_{\bf k}^1\\ \cdots \\  
 ~E_{\bf k}^{N_l}\\ 
 \end{array} \right). 
\end{eqnarray} 
Next, we use the following relation (see appendix \ref{app:G}) 
\begin{eqnarray}
 && 
  \left( \begin{array}{llll} 
  a^{(l)}_{00} & a^{(l)}_{10}  & \ldots & a^{(l)}_{N_l, 0}  \\ 
  0 & a^{(l)}_{11}  & \ldots & a^{(l)}_{N_l,1} \\ 
 \ldots        &\ldots & \ldots  &\ldots \\  
 0  &  0 &\ldots &  a^{(l)}_{N_l,N_l} \\ 
 \end{array} \right) 
  \left( \begin{array}{lllll} 
  a^{(l)}_{00} &0& \ldots & 0  \\ 
  a^{(l)}_{10} & a^{(l)}_{11}  & \ldots   & 0 \\ 
 \ldots     &\ldots     &\ldots   & 0 \\  
 a^{(l)}_{N_l,0}    & a^{(l)}_{N_l,1} & \ldots &  a^{(l)}_{N_l,N_l} \\ 
 \end{array} \right) \nonumber \\ 
&&\qquad = 
 \left( \begin{array}{llll} 
  I^{(l)}_{0} & I^{(l)}_{1}  & \ldots & I^{(l)}_{N_l}  \\ 
  I^{(l)}_{1} & I^{(l)}_{2}  & \ldots & I^{(l)}_{N_l+1} \\ 
 \ldots        &\ldots & \ldots  &\ldots \\  
 I^{(l)}_{N_l}  & I^{(l)}_{N_l+1} &\ldots &  I^{(l)}_{2N_l} \\ 
 \end{array} \right)^{-1}, 
 \label{eq:orthogonal_in_matrix}
\end{eqnarray}
which can be proved by the orthogonal condition 
expressed by eq.(\ref{eq:orthogonal_matrix}), 
\begin{eqnarray}	
 \sum_{k=0}^{n} I^{(l)}_{r+k} a^{(l)}_{nk} = \delta_{rn}\frac{1}{a_{nn}} . 
 \label{eq:orthogonal_in_components}
\end{eqnarray}  
Hence, we obtain 
\begin{eqnarray}
\lambda_{\bf k}^{\langle \mu_1\cdots \mu_l \rangle} 
&=&\frac{{\cal W}^{(l)}}{l!}\left(
  \begin{array}{llll} 
  \rho_0^{\mu_1\cdots\mu_l},&
  \rho_1^{\mu_1\cdots\mu_l},&
  \cdots, &
  \rho_{N_l}^{\mu_1\cdots\mu_l}
  \end{array} 
   \right) \nonumber \\ 
&& \qquad \qquad \times  
\left( \begin{array}{llll} 
  I^{(l)}_{0} & I^{(l)}_{1}  & \ldots & I^{(l)}_{N_l}  \\ 
  I^{(l)}_{1} & I^{(l)}_{2}  & \ldots & I^{(l)}_{N_l+1} \\ 
 \ldots        &\ldots & \ldots  &\ldots \\  
 I^{(l)}_{N_l}  & I^{(l)}_{N_l+1} &\ldots &  I^{(l)}_{2N_l} \\ 
 \end{array} \right)^{-1} 
 \left( \begin{array}{c} 
 E_{\bf k}^0 \\ 
 E_{\bf k}^1\\ \cdots \\  
~E_{\bf k}^{N_l}\\ 
 \end{array} \right).  
\label{eq:lambda_matrix2} 
\end{eqnarray}
By integrating in whole momentum space with multiplying 
the weight factor $\omega_{\bf k}^{(l)}$ for the 
both sides of eq.(\ref{eq:lambda_matrix2}), 
one obtain 
\begin{eqnarray}
 \frac{l!}{{\cal W}^{(l)}} &&\hspace*{-8mm} 
   \int\!\! \frac{d^3{\bf k}~\omega^{(l)}_{\bf k}}{(2\pi)^3} 
   \lambda_{\bf k}^{\langle \mu_1\cdots\mu_l\rangle} E_{\bf k}^r 
   \nonumber \\ 
 &=& \left(
  \begin{array}{llll} 
  \rho_0^{\mu_1\cdots\mu_l},&
  \rho_1^{\mu_1\cdots\mu_l},&
  \cdots, &
  \rho_{N_l}^{\mu_1\cdots\mu_l}
  \end{array} 
   \right)  
 \left( \begin{array}{llll} 
  I^{(l)}_{0} & I^{(l)}_{1}  & \ldots & I^{(l)}_{N_l}  \\ 
  I^{(l)}_{1} & I^{(l)}_{2}  & \ldots & I^{(l)}_{N_l+1} \\ 
 \ldots        &\ldots & \ldots  &\ldots \\  
 I^{(l)}_{N_l}  & I^{(l)}_{N_l+1} &\ldots &  I^{(l)}_{2N_l} \\ 
 \end{array} \right)^{-1} 
 \left(\begin{array}{l}
  I^{(l)}_{r} \\ I^{(l)}_{1+r}  \\ \ldots \\ I^{(l)}_{N_l+r}  \\ 
 \end{array}\right) \nonumber \\ 
&=&  \rho_r^{\mu_1\cdots\mu_l}, \label{eq:addition_31}
\end{eqnarray}
because of the characteristics of the inverse matrix in eq.(\ref{eq:addition_31}). 
Thus we finally obtain
\begin{eqnarray}
    \rho_r^{\mu_1\cdots\mu_l}
&=&  \! \frac{l!}{{\cal W}^{(l)}}
   \int\!\! \frac{d^3{\bf k}~\omega^{(l)}_{\bf k}}{(2\pi)^3} 
   \lambda_{\bf k}^{\langle \mu_1\cdots\mu_l\rangle} E_{\bf k}^r,  
    \nonumber \\ 
&=& \frac{l!}{(2l+1)!!} \! 
   \int\!  \frac{d^3{\bf k}}{(2\pi)^3} 
   \lambda_{\bf k}^{\langle \mu_1\cdots\mu_l\rangle}
   [\Delta^{\alpha\beta}k_{\alpha}k_{\beta}]^l 
   E_{\bf k}^r f_{{\bf k}0}, 
\end{eqnarray}
which is eq.(\ref{eq:from_micro_to_macro1}). 
\section{Polynomial $P_{{\bf k}n}^{(m)}$ and its coefficients $a_{nr}^{(l)}$}\label{app:G}
The orthogonal condition of the polynomial $P_{{\bf k}n}^{(m)}$ in 
eq. (\ref{eq:orthogonality_conditionPn_(a)}) can be written 
in the following form 
\begin{eqnarray}
  \left( \begin{array}{llllll}
  1   & I^{(l)}_1 && \ldots &&  I^{(l)}_{n}    \\
  I^{(l)}_1 & I^{(l)}_2 && \ldots &&  I^{(l)}_{n+1} \\ 
  \hdotsfor{6} \\ 
 I^{(l)}_{n-1} & I^{(l)}_{n} && \ldots && I^{(l)}_{2n-1} \\ 
 I^{(l)}_n & I^{(l)}_{n+1} && \ldots && I^{(l)}_{2n} \\
\end{array}   \right) 
\left( \begin{array}{c} 
 a^{(l)}_{n0} \\  a^{(l)}_{n1} \\ \cdots \\  a^{(l)}_{n n-1} \\  
 a^{(l)}_{nn} \\ 
 \end{array} \right)  = 
\left( \begin{array}{c} 
 0 \\  0\\ \cdots \\  0 \\  
 1/a^{(l)}_{nn} \\ 
 \end{array} \right), 
\end{eqnarray} 
and 
\begin{eqnarray}
    a_{nk}^{(l)}= \frac{{\cal I}^{(l)}_{nk}}{\sqrt{{\rm det}I_n^{(l)} {\cal I}_{nn}^{(l)}
}},
\end{eqnarray}
where ${\rm det}I_n^{(l)}$ is the determinant of the matrix
\begin{eqnarray}
   \{ I_n^{(l)}\}_{ij} = \left( \begin{array}{llllll}
  1   & I^{(l)}_1 && \ldots &&  I^{(l)}_{n}    \\
  I^{(l)}_1 & I^{(l)}_2 && \ldots &&  I^{(l)}_{n+1} \\ 
  \hdotsfor{6} \\ 
 I^{(l)}_{n-1} & I^{(l)}_{n} && \ldots && I^{(l)}_{2n-1} \\ 
 I^{(l)}_n & I^{(l)}_{n+1} && \ldots && I^{(l)}_{2n} \\
\end{array}   \right), 
\end{eqnarray}
and ${\cal I}_{nk}^{(l)}$ is a co-factor (cross out the entries that lie in 
the corresponding row $n$ and column $k$) of the matrix $I_n^{(l)}$.   
Thus, we can express the orthogonal function as the following
\begin{eqnarray}
    P_{{\bf k}n}^{(l)}= \sum_{r=0}^{N_l} 
    \frac{{\cal I}^{(l)}_{nr} E_{\bf k}^r  }{\sqrt{{\rm det}I_n^{(l)} {\cal I}_{nn}^{(l)}}}.
    \label{eq:anm} 
\end{eqnarray}
Then for eq.(\ref{eq:orthogonal_in_matrix}) we obtain, for example, in the
 $N_l=1$ case 
\begin{eqnarray} 
 \left(\begin{array}{cc}
 a_{00}^{(l)} & a_{10}^{(l)} \\ 
   0     & a_{11}
 \end{array}\right)
 \left(\begin{array}{cc}
 a_{00}^{(l)} & 0      \\ 
 a_{10}^{(l)} & a_{11}^{(l)}
 \end{array}\right) = 
\left(\begin{array}{cc}
 I_0^{(l)} & I_1^{(l)}      \\ 
 I_1^{(l)} & I_2^{(l)}
 \end{array}\right)^{-1}.
\end{eqnarray}  
Therefore,  
\begin{eqnarray}
 \left(\begin{array}{cc}
 a_{00}^{(l)} & a_{10}^{(l)} \\ 
   0     & a_{11}^{(l)}
 \end{array}\right)
 \left(\begin{array}{cc}
 a_{00}^{(l)} & 0      \\ 
 a_{10}^{(l)} & a_{11}^{(l)}
 \end{array}\right)
\left(\begin{array}{cc}
 I_0^{(l)} & I_1^{(l)}      \\ 
 I_1^{(l)} & I_2^{(l)}
 \end{array}\right) 
=  \left(\begin{array}{cc}
 1 & 0 \\ 
   0     & 1
 \end{array}\right). \qquad
\end{eqnarray}  
This is satisfied if eq.(\ref{eq:orthogonal_in_components}) holds true; 
i.e., for the orthogonal $0,0$ and $1,1$ components, respectively
\begin{subequations} 
\begin{eqnarray}
&& ((a_{00}^{(l)})^2+(a_{10}^{(l)})^2)I_0^{(l)}+ a_{10}^{(l)}a_{11}^{(l)} 
I_1^{(l)} =1 +a_{10}(I_0a_{10}+I_1a_{11}) \quad =1,\\ 
&& a_{10}^{(l)}a_{11}^{(l)}I_1^{(l)}+(a_{11}^{(l)})^2I_2^{(l)} = a_{11}^{(l)}(I_1^{(l)}a_{10}^{(l)}+I_2a_{11}^{(l)})\quad =1, 
\end{eqnarray}
where we use $a_{00}^{(l)}=1$ and $I_0^{(l)}=1$. 
However, the off-orthogonal ($0,1$) and ($1,0$) components are
\begin{eqnarray}
&& ((a_{00}^{(l)})^2+(a_{10}^{(l)})^2)I_1+a_{10}^{(l)}a_{11}^{(l)}I_2^{(l)} \nonumber \\ 
&&~= I_1^{(l)} +a_{10}^{(l)}(I_1^{(l)}a_{10}^{(l)}+I_2^{(l)} a_{11}^{(l)})=
I_1^{(l)}+a_{10}^{(l)}/a_{11}^{(l)}  \nonumber \\ 
&&~= (I_0^{(l)}a_{10}^{(l)}+I_1^{(l)}a_{11}^{(l)})/a_{11}^{(l)}=0, \\ 
&& a_{10}^{(l)}a_{11}^{(l)}I_0^{(l)} + (a_{11}^{(l)})^2I_1^{(l)}= 
a_{11}^{(l)}(I_0^{(l)}a_{10}^{(l)}+I_1^{(l)}a_{11}^{(l)}) \quad =0, 
\end{eqnarray}
\end{subequations} 
respectively. 

\bibliographystyle{ptephy}
\bibliography{separ15}
\end{document}